\documentclass[usenatbib]{mnras}
\usepackage{graphicx}
\usepackage{dcolumn}
\usepackage{amssymb} 
\usepackage{amsmath}
\usepackage{ctable}
\usepackage{float}
\usepackage{bm}
\usepackage{float}
\usepackage{epsfig}
\usepackage{epstopdf}
\usepackage{epsf,color}
\usepackage{comment}
\usepackage{aas_macros}
\usepackage{url}
\usepackage{soul}
\usepackage{color, colortbl}
\usepackage{widetext}
\usepackage[T1]{fontenc}

\newcommand{\cmnt}[1]{}

\makeatletter 
  \patchcmd{\NAT@citex}
    {\@citea\NAT@hyper@{%
      \NAT@nmfmt{\NAT@nm}%
      \hyper@natlinkbreak{\NAT@aysep\NAT@spacechar}{\@citeb\@extra@b@citeb}%
      \NAT@date}}
    {\@citea\NAT@nmfmt{\NAT@nm}%
    \NAT@aysep\NAT@spacechar\NAT@hyper@{\NAT@date}}{}{}

  \patchcmd{\NAT@citex}
    {\@citea\NAT@hyper@{%
      \NAT@nmfmt{\NAT@nm}%
      \hyper@natlinkbreak{\NAT@spacechar\NAT@@open\if*#1*\else#1\NAT@spacechar\fi}%
        {\@citeb\@extra@b@citeb}%
      \NAT@date}}
    {\@citea\NAT@nmfmt{\NAT@nm}%
    \NAT@spacechar\NAT@@open\if*#1*\else#1\NAT@spacechar\fi\NAT@hyper@{\NAT@date}}
    {}{}
\makeatother

\usepackage[normalem]{ulem}
\usepackage{color}

\usepackage{tgtermes}
\newcommand{\hii}{\ion{H}{ii}}
\newcommand{\hi}{\ion{H}{i}}

\newcommand{\oi}{\ion{O}{i}}
\newcommand{\oiii}{\ion{O}{iii}}
\newcommand{\oii}{\ion{O}{ii}}

\newcommand{\nii}{\ion{N}{ii}}
\newcommand{\niii}{\ion{N}{iii}}

\newcommand{\sii}{\ion{S}{ii}}

\title[Strong line diagnostics]{Analytical strong line diagnostics and their redshift evolution}

\author[S. Yang et al.]{%
Shengqi Yang,$^{1,2}$\thanks{E-mail: \href{mailto:shengqiy@lanl.gov}{shengqiy@lanl.gov}}
Adam Lidz,$^{3}$
Andrew Benson,$^{1}$
Swathya Singh Chauhan,$^{4}$
Aaron Smith$^{5}$
\newauthor
and
Hui Li$^{6}$
\\
$^{1}$Carnegie Observatories, 813 Santa Barbara Street, Pasadena, CA 91101, USA\\
$^{2}$Los Alamos National Laboratory, NM 87545, USA\\
$^{3}$Department of Physics and Astronomy, University of Pennsylvania, 209 South 33rd Street, Philadelphia, PA 19104, USA\\
$^{4}$Department of Physics and Astronomy, University of California, Los Angeles, CA, 90095, USA\\
$^{5}$Department of Physics, The University of Texas at Dallas, Richardson, Texas 75080, USA\\
$^{6}$Department of Astronomy, Tsinghua University, Beijing 100084, China\\
}

\date{Accepted XXX. Received YYY; in original form ZZZ}

\pubyear{2023}

\begin{document}
\label{firstpage}
\pagerange{\pageref{firstpage}--\pageref{lastpage}}
\maketitle

\begin{abstract}
The \textit{JWST} is allowing new measurements of gas-phase metallicities in galaxies between cosmic noon and cosmic dawn. The most robust approach uses luminosity ratios between the excited auroral transition, [\oiii] 4364\,\AA, and the lower [\oiii] 5008\,\AA/4960\,\AA\ lines to determine the gas temperature. The ratio of the luminosities in the latter transitions to those in hydrogen Balmer series lines then yield relatively clean metallicity estimates. In the absence of detection of the [\oiii] auroral line, the ratios of various [\oiii], [\oii], [\nii], and Balmer lines are used to determine metallicities. Here we present a refined approach for extracting metallicities from these ``strong line diagnostics''. Our method exploits empirical correlations between the temperature of \oiii/\oii\ regions and gas-phase metallicity. We then show, from first principles, how to extract metallicities  and break degeneracies in these estimates using traditional strong line diagnostics, R2, R3, R23, and O3O2, and N2O2. We show that these ratios depend also on volume correction factors, i.e. on accounting for the fraction of the volume of HII regions that are in \oiii\ and \oii, but that these can be determined self-consistently along with the metallicities. We quantify the success of our method using metallicities derived from galaxies with auroral line determinations and show that it generally works better than previous empirical approaches. 
The scatter in the observed line ratios and redshift evolution are largely explained by O3O2 variations. 
We provide publicly available routines for extracting metallicities from strong line diagnostics using our methodology.
\end{abstract}




\section{Introduction}
One key to tracing and understanding galaxy formation and evolution is to determine the cosmic chemical enrichment history. An important goal here is to robustly determine the average redshift evolution of the gas-phase metallicity in the interstellar medium (ISM) of distant galaxies. In addition, there is a ``mass--metallicity relationship'', expressing correlations between the gas-phase metallicity and stellar mass (e.g. \citealt{2004ApJ...613..898T,2006ApJ...647..970L,2008A&A...488..463M,2009MNRAS.398.1915M,2013ApJ...771L..19Z,2021ApJ...919..143H}). This is thought to be controlled, in part, by feedback processes \citep{2016MNRAS.456.2140M,2023MNRAS.518.3557U}; combining measurements of metallicity and stellar mass across cosmic time may help in understanding the uncertain role of feedback in self-regulating galaxy formation. Among the important probes of ISM metallicity are rest-frame optical emission lines from a broad range of heavy elements, including [OIII], [OII], and [NII] collisional exictation lines. Although these lines, when emitted from $z \gtrsim 3$, are generally unobservable from the ground owing to absorption in the Earth's atmosphere, infra-red measurements from the \textit{James Webb Space Telescope} (\textit{JWST}) are enabling rapid progress, including detections into the Epoch of Reionization \citep[EoR; e.g.][]{2023NatAs.tmp..194H,2023MNRAS.518..425C,2024ApJ...962...24S,2023arXiv230603120L}.

A technique known as the ``direct $T_e$ method'' provides one of the best approaches for determining the ISM metallicity from \textit{JWST} observations (e.g. \citealt{2017PASP..129h2001P,2019A&ARv..27....3M}). This
method exploits measurements of multiple rest-frame optical [\oiii] and [\oii] lines
in combination with hydrogen Balmer line observations. The relevant lines
have high critical densities such that their luminosities are insensitive to the \hii\ region gas densities, and instead depend mainly on ISM metallicity and gas temperature. In the direct $T_e$ method, the gas temperature is determined via the luminosity ratios between auroral [\oiii] $\lambda 4364$\,\AA\ lines and rest-frame optical [\oiii] transitions between lower energy states. The metallicity then follows from the temperature determination
and measurements of luminosity ratios between [\oiii], [\oii], and hydrogen Balmer lines. The main disadvantage of this technique is that the
[\oiii] $\lambda 4364$\,\AA\ line is often 10--100 times fainter than the nearby [\oiii] $\lambda 5008$\,\AA, [\oiii] $\lambda 4960$\,\AA, and H$\beta$ lines. 
It is therefore useful to develop ``strong line diagnostic'' ratios which aim to provide robust metallicity estimates in the absence of [\oiii] $\lambda 4364$~\AA\ detections. In particular, previous efforts have used samples of galaxies with direct $T_e$ metallicity measurements to calibrate metallicity estimates from the strong-line observations alone. These have included
purely empirical approaches (e.g. \citealt{2004MNRAS.348L..59P,2005ApJ...631..231P,2008A&A...488..463M,2010ApJ...720.1738P,2015ApJ...813..126J,2017MNRAS.465.1384C,2018ApJ...859..175B}),
work employing numerical spectral synthesis codes (such as \textsc{Cloudy} and similar tools, e.g. \citealt{1994ApJ...420...87Z,1991ApJ...380..140M,2002ApJS..142...35K,2004ApJ...617..240K,2011A&A...526A.149N,2016Ap&SS.361...61D}), and studies which combine empirical and theoretical aspects (e.g. \citealt{2002MNRAS.330...69D,2006A&A...459...85N,2008A&A...488..463M}). A detailed review of different strong line diagnostic methods is given in
\cite{2019A&ARv..27....3M}.
One potential concern here is that empirical relationships of strong line diagnostic are often calibrated using only low redshift samples, and the same relationships may not be applicable at higher redshifts \citep{2023arXiv230603120L,2023arXiv230503753H}.\par
In this work, we consider strong line diagnostics from a first-principles perspective. Specifically, we derive expressions for strong line diagnostics assuming: (\textit{i}) spherical and ionization-bounded \hii\ regions, (\textit{ii})
all relevant lines are in the low-density limit so that collisional de-excitations may be ignored, (\textit{iii}) the oxygen in each \hii\ region is predominantly in the form of \oii\ or \oiii, (\textit{iv}) tight correlations exist between gas-phase metallicities and the gas temperatures in each of the \oii\ and \oiii\ regions (\citealt{2023MNRAS.tmp.2479Y}; hereafter \defcitealias{2023MNRAS.tmp.2479Y}{Y23}\citetalias{2023MNRAS.tmp.2479Y}),
and (\textit{v}) gas temperature fluctuation within each galaxy is negligible.   The assumed relation between gas temperature and gas-phase metallicity (referred to as the ``TZR'') effectively eliminates a degree-of-freedom from the problem and allows more transparent metallicity determinations using strong line diagnostics. While the TZRs are calibrated using direct $T_e$ samples, which brings an empirical element into our modeling, we show that the empirical TZRs are close to theoretical \textsc{Cloudy} thermal equilibrium calculations. Under these simplifying assumptions, following 
recent studies (\cite{2020MNRAS.499.3417Y}; \defcitealias{2023MNRAS.tmp.2479Y}{Y23}\citetalias{2023MNRAS.tmp.2479Y}), we derive
analytic expressions for multiple strong line diagnostic ratios. 

In this work, we focus on the following luminosity ratios among strong [\oiii], [\oii], and [\nii] lines. These are relevant for current high-redshift metallicity measurements from the \textit{JWST}:
\begin{equation}\label{eq:StrongLineDiagnostics}
\begin{split}
    \mathrm{R2}&=L^\mathrm{[\oii]}_{3727,30}/L_\mathrm{H\beta}\,,\\
    \mathrm{R3}&=L^\mathrm{[\oiii]}_{5008}/L_\mathrm{H\beta}\,,\\
    \mathrm{R23}&=(L^\mathrm{[\oii]}_{3727,30}+L^\mathrm{[\oiii]}_{4960}+L^\mathrm{[\oiii]}_{5008})/L_\mathrm{H\beta}\,,\\
    \mathrm{O3O2}&=L^\mathrm{[\oiii]}_{5008}/L^\mathrm{[\oii]}_{3727,30}\,,\\
    \mathrm{N2O2}&=L^{[\nii]}_{6584}/L^\mathrm{[\oii]}_{3727,30}\,.
\end{split}
\end{equation}\par 
In Section~\ref{sec:diagnostics}, we derive expressions for these strong line diagnostic ratios under the above assumptions, following related calculations in \citet{2020MNRAS.499.3417Y} and \defcitealias{2023MNRAS.tmp.2479Y}{Y23}\citetalias{2023MNRAS.tmp.2479Y}. Section~\ref{sec:directTe} discusses the direct $T_e$ measurement data used to in this work to calibrate and test our technique.  
Next, we show that R2, R3, and R23 depend not only on the metallicity of the \hii\ regions, $Z$, but also on
volume correction factors (VCFs) which characterize the fraction of the volume of the \hii\ regions that are in each of \oii\ and \oiii. The VCFs, however, can be extracted from the O3O2 diagnostic. The VCFs are themselves interesting because they provide information regarding the spectral shape of the ionizing radiation in the observed galaxies. We consider the implications of current O3O2 measurements in Section~\ref{sec:ICFs}, while
in Section~\ref{sec:Tgas} we use the current direct $T_e$ samples to extract empirical correlations between \oiii\ and \oii\ region gas temperatures and gas-phase metallicities. These are compared to theoretical model calculations which assume thermal equilibrium. We also suggest that the N2O2 diagnostic may help in breaking degeneracies among the multiple metallicity solutions allowed by R2, R3, and R23 alone. We then infer metallicities from our strong line diagnostic relations and compare with those extracted from the direct $T_e$ technique in Section~\ref{sec:Obs} (the direct $T_e$ estimates are assumed to capture the true metallicities). We
contrast these results with other strong line diagnostic methods in the literature. We show that our method generally performs better than previous determinations in the literature, while our results may be further refined in the future with improved gas temperature models and/or calibrations. We attribute much of the scatter and redshift evolution in the strong line diagnostic metallicity estimates to variations in the shape of the ionizing radiation spectrum. Finally, we summarize the assumptions, strengths, and limitations of our calculations in Section~\ref{sec:summary}. We provide
a publicly available code which implements our approach for
constraining metallicites from R2, R3, R23, O3O2, and N2O2 measurements at \url{https://github.com/Sheng-Qi-Yang/HIILines}.

\section{Strong line diagnostics}\label{sec:diagnostics}

To start, we can consider the simplified one-zone picture in which a single radiation source ionizes a surrounding spherical region with uniform ISM properties. Further, we suppose that the gas density in the ionized ISM is much lower than the critical densities for the emission lines of interest. In this toy scenario, the [\oiii], [\oii], H$\beta$, and [\nii] line luminosities (Eqs. 11 and 15 of \defcitealias{2023MNRAS.tmp.2479Y}{Y23}\citetalias{2023MNRAS.tmp.2479Y}) follow:
\begin{equation}
    L^\mathrm{[\oii]}_{3727,30}=\left(\dfrac{n_\mathrm{O}}{n_\mathrm{H}}\right)_\odot\dfrac{Z}{Z_\odot}h(k_{01}^\mathrm{\oii}\nu_{10}^\mathrm{\oii}+k_{02}^\mathrm{\oii}\nu_{20}^\mathrm{\oii})\dfrac{Q_\mathrm{\hi}}{\alpha_\mathrm{B,\hii}}\dfrac{V_\mathrm{\oii}}{V_\mathrm{\hii}}\,,
\end{equation}
\begin{equation}
    L^\mathrm{[\oiii]}_{5008}=\left(\dfrac{n_\mathrm{O}}{n_\mathrm{H}}\right)_\odot\dfrac{Z}{Z_\odot}\dfrac{A_{32}^\mathrm{\oiii}k_{03}^\mathrm{\oiii}h\nu_{32}^\mathrm{\oiii}}{A_{31}^\mathrm{\oiii}+A_{32}^\mathrm{\oiii}}\dfrac{Q_\mathrm{\hi}}{\alpha_\mathrm{B,\hii}}\dfrac{V_\mathrm{\oiii}}{V_\mathrm{\hii}}\,,
\end{equation}
\begin{equation}
    L^\mathrm{[\oiii]}_{4960}=\left(\dfrac{n_\mathrm{O}}{n_\mathrm{H}}\right)_\odot\dfrac{Z}{Z_\odot}\dfrac{A_{31}^\mathrm{\oiii}k_{03}^\mathrm{\oiii}h\nu_{31}^\mathrm{\oiii}}{A_{31}^\mathrm{\oiii}+A_{32}^\mathrm{\oiii}}\dfrac{Q_\mathrm{\hi}}{\alpha_\mathrm{B,\hii}}\dfrac{V_\mathrm{\oiii}}{V_\mathrm{\hii}}\,,
\end{equation}
\begin{equation}
    L^\mathrm{[\nii]}_{6584}=\left(\dfrac{n_\mathrm{N}}{n_\mathrm{H}}\right)_\odot\dfrac{Z}{Z_\odot}\dfrac{A_{32}^\mathrm{\nii}k_{03}^\mathrm{\nii}h\nu_{32}^\mathrm{\nii}}{A_{31}^\mathrm{\nii}+A_{32}^\mathrm{\nii}}\dfrac{Q_\mathrm{\hi}}{\alpha_\mathrm{B,\hii}}\dfrac{V_\mathrm{\nii}}{V_\mathrm{\hii}}\,,
\end{equation}
\begin{equation}
    L_\mathrm{H\beta}=h\nu_\mathrm{H\beta}\dfrac{\alpha_\mathrm{B,H\beta}}{\alpha_\mathrm{B,\hii}}Q_\mathrm{\hi}\,.
\end{equation}
Here $(n_\mathrm{O}/n_\mathrm{H})_\odot=10^{-3.31}$ and $(n_\mathrm{N}/n_\mathrm{H})_\odot=10^{-4.07}$ are the oxygen and nitrogen versus hydrogen number density ratios at Solar metallicity \citep{2001AIPC..598...23H,2001ApJ...556L..63A}; $h$ is Planck's constant; $Q_\mathrm{\hi}$ is the rate of production of hydrogen ionizing photons;
$V_\mathrm{\oii}$, $V_\mathrm{\oiii}$, $V_\nii$, and $V_\mathrm{\hii}$ 
are, respectively, the volumes of the \oii, \oiii, \nii, and \hii\ regions around the ionizing source; 
$A_{ij}^X$, $k_{ij}^X$, and $\nu_{ij}^X$ are the spontaneous decay rate, collisional excitation rate, and the rest-frame emission frequency for ion $X$ transitioning from energy level $i$ to level $j$. 
Further,
$\alpha_\mathrm{B,\hii}$ is the case B \hii\ recombination rate, while $\alpha_\mathrm{B,H\beta}/\alpha_\mathrm{B,\hii}$ is the fraction of hydrogen recombinations that result in H$\beta$ line emission. 
For simplicity, in what follows we assume that the N/O ratio for all galaxies follows the Solar abundance ratio. In practice, low metallicity and/or high redshift galaxies likely depart from this assumption \citep{2014AJ....147..131P}: in future applications, one may want to modify the value adopted for the N/O ratio. 

We can then generalize to the more realistic case where a galaxy contains numerous \hii\ regions, each with a unique hydrogen ionizing photon production rate, $Q_\mathrm{\hi}$, gas-phase metallicity, $Z$, gas density, $n_\mathrm{H}$, and VCFs, $V_\mathrm{\oii}/V_\mathrm{\hii}$, $V_\mathrm{\oiii}/V_\mathrm{\hii}$, $V_\mathrm{\nii}/V_\mathrm{\hii}$. 
As mentioned previously, we assume that each \hii\ region is ionization-bounded so that the entire supply of ionizing photons is absorbed locally. In this case, the resulting galaxy-wide total line luminosities can be expressed as: 
\begin{equation}\label{eq:multiZone}
\begin{split}
    L^\mathrm{[\oii]}_{3727,30}=&\left(\dfrac{n_\mathrm{O}}{n_\mathrm{H}}\right)_\odot \dfrac{h(k_{01}^\mathrm{\oii}\nu_{10}^\mathrm{\oii}+k_{02}^\mathrm{\oii}\nu_{20}^\mathrm{\oii})}{\alpha_\mathrm{B,\hii}}\\
    &\times\sum Q_\mathrm{\hi}\dfrac{Z}{Z_\odot}\dfrac{V_\mathrm{\oii}}{V_\mathrm{\hii}}\,,
\end{split}
\end{equation}
\begin{equation}
    L^\mathrm{[\oiii]}_{5008}=\dfrac{3}{4}\left(\dfrac{n_\mathrm{O}}{n_\mathrm{H}}\right)_\odot \dfrac{k_{03}^\mathrm{\oiii}h\nu_{32}^\mathrm{\oiii}}{\alpha_\mathrm{B,\hii}}\times\sum Q_\mathrm{\hi}\dfrac{Z}{Z_\odot}\dfrac{V_\mathrm{\oiii}}{V_\mathrm{\hii}}\,,
\end{equation}
\begin{equation}
    L^\mathrm{[\oiii]}_{4960}=\dfrac{1}{4}\left(\dfrac{n_\mathrm{O}}{n_\mathrm{H}}\right)_\odot\dfrac{k_{03}^\mathrm{\oiii}h\nu_{31}^\mathrm{\oiii}}{\alpha_\mathrm{B,\hii}}\times\sum Q_\mathrm{\hi}\dfrac{Z}{Z_\odot}\dfrac{V_\mathrm{\oiii}}{V_\mathrm{\hii}}\,,
\end{equation}
\begin{equation}
    L^\mathrm{[\nii]}_{6584}=\dfrac{3}{4}\left(\dfrac{n_\mathrm{N}}{n_\mathrm{H}}\right)_\odot \dfrac{k_{03}^\mathrm{\nii}h\nu_{32}^\mathrm{\nii}}{\alpha_\mathrm{B,\hii}}\times\sum Q_\mathrm{\hi}\dfrac{Z}{Z_\odot}\dfrac{V_\mathrm{\nii}}{V_\mathrm{\hii}}\,,
\end{equation}
\begin{equation}
\label{eq:hbeta_multiz}
    L_\mathrm{H\beta}=h\nu_\mathrm{H\beta}\dfrac{\alpha_\mathrm{B,H\beta}}{\alpha_\mathrm{B,\hii}}\times\sum Q_\mathrm{\hi}\,,
\end{equation}
where the $\sum$ sums over all of the \hii\ regions within a galaxy. Here we have used the fact that $A_{32}^\mathrm{\oiii}=3A_{31}^\mathrm{\oiii}$ and $A_{32}^\mathrm{\nii}=3A_{31}^\mathrm{\nii}$. 

Note that our model describes zones of \oiii, \oii, and \nii\ within each \hii\ region and properly allows for differences in the temperature of these zones (see \S~\ref{sec:Tgas}), e.g. the \oiii\ and \oii\ zones generally have different temperatures. However, an important simplifying assumption in this work is to ignore scatter in the temperature across each galaxy and to further adopt a uniform temperature within each zone. That is, the different \oiii\ regions in a given galaxy are assumed to have a common and uniform temperature, while a similar assumption applies to \oii\ regions, etc. In this case, each galaxy is characterized by four related temperatures: $T_4^\mathrm{\oiii}$, $T_4^\mathrm{\oii}$, $T_4^\mathrm{\nii}$, $T_4^\mathrm{\hii}$,
where $T_4^X$ denotes the gas temperature, in units of $10^4$ K, for regions predominantly containing the ion denoted by X.
The uniform temperature
assumption allows us to pull the temperature dependent collisional excitation and radiative recombination coefficients (i.e. $k_{ij}^X$, $\alpha_\mathrm{B,\hii}$, and $\alpha_\mathrm{B,H\beta}$) out of the summation, as in Eqs.~(\ref{eq:multiZone})--(\ref{eq:hbeta_multiz}). 
Although we ignore temperature fluctuations in this work, this approximation is also inherent to the direct $T_e$ method itself, at least as commonly implemented. 
This simplification does receive some observational support: specifically, \cite{2023NatAs.tmp...87C} compares the gas-phase metallicities in Mrk71 extracted from temperature insensitive [\oiii] 88, 52 $\mu$m lines and that from temperature sensitive [\oiii] 5008, 4960, 4364~\AA\ lines. Those authors use the agreement between the metallicities from these two approaches to bound the level of temperature fluctuations across \oiii\ regions to $\lesssim 3,000$ K. On the other hand, the evidence for this uniform temperature approximation is limited and controversial (e.g. \citealt{2023Natur.618..249M}). Furthermore, note that metallicity gradients are commonly observed in galaxies and that these may lead to cooling-rate and hence temperature variations across each galaxy, although the quantitative impact here is unclear (e.g. \cite{2009ApJ...700..309B,2013ApJ...777...96C,2015AJ....149..107J,2021ApJ...923..203S,2024arXiv240303977V}). We believe the uniform temperature approximation is a plausible starting assumption. 

We can then write direct expressions for the strong line diagnostics as:
\begin{widetext}
\begin{equation}\label{eq:diagnostic_T}
    \begin{split}
        \log(\mathrm{R2})=&-3.19+\log\left[\dfrac{k_{01}^\mathrm{\oii}(T_4^\mathrm{\oii})+k_{02}(T_4^\mathrm{\oii})}{\alpha_\mathrm{B,H\beta}(T_4^\mathrm{\hii})}\right]+\log(Z_\mathrm{Q})+\log(\mathrm{VCF}_\mathrm{\oii})\,,\\
        \log(\mathrm{R3})=&-3.45+\log\left[\dfrac{k_{03}^\mathrm{\oiii}(T_4^\mathrm{\oiii})}{\alpha_\mathrm{B,H\beta}(T_4^\mathrm{\hii})}\right]+\log(Z_\mathrm{Q})+\log(\mathrm{VCF}_\mathrm{\oiii})\,,\\
        \log(\mathrm{R23})=&-3.31+\log\left[\dfrac{1.30[k_{01}^\mathrm{\oii}(T_4^\mathrm{\oii})+k_{02}^\mathrm{\oii}(T_4^\mathrm{\oii})]\mathrm{VCF}_\oii}{\alpha_\mathrm{B,H\beta}(T_4^\mathrm{\hii})}+\dfrac{0.973k_{03}^\mathrm{\oiii}(T_4^\mathrm{\oiii})\mathrm{VCF}_\oiii}{\alpha_\mathrm{B,H\beta}(T_4^\mathrm{\hii})}\right]+\log(Z_\mathrm{Q})\,,\\
        \log(\mathrm{O3O2})=&\log\left[\dfrac{0.559k_{03}^\mathrm{\oiii}(T_4^\mathrm{\oiii})}{k_{01}^\mathrm{\oii}(T_4^\mathrm{\oii})+k_{02}^\mathrm{\oii}(T_4^\mathrm{\oii})}\right]+\log(\mathrm{VCF}_\oiii)-\log(\mathrm{VCF}_\oii)\,,\\
        \log(\mathrm{N2O2})=&\log\left[\dfrac{0.0738k_{03}^\mathrm{\nii}(T_4^\mathrm{\nii})}{k_{01}^\mathrm{\oii}(T_4^\mathrm{\oii})+k_{02}^\mathrm{\oii}(T_4^\mathrm{\oii})}\right]+\log(\mathrm{VCF}_\mathrm{\nii})-\log(\mathrm{VCF}_\mathrm{\oii})\,,\\
    \end{split}
\end{equation}
\end{widetext}
The expressions above imply that the metallicity accessed by the strong line diagnostics is a $Q_\mathrm{\hi}$-weighted metallicity averaged across the \hii\ regions of the measured galaxy (\defcitealias{2023MNRAS.tmp.2479Y}{Y23}\citetalias{2023MNRAS.tmp.2479Y}):
\begin{equation}\label{eq:ZQ}
    Z_\mathrm{Q}=(\Sigma Q_\mathrm{\hi}Z)/(\Sigma Q_\mathrm{\hi})\,.
\end{equation}
In addition to the dependence on this ``effective \hii\ region metallicity'', however, the R2 and R3 diagnostics also depend on the VCFs. Specifically, 
we have introduced the quantities $\mathrm{VCF}_{\oii}$ and $\mathrm{VCF}_{\oiii}$ which depend on how much of the \hii\ region volume is in each of \oii\ and \oiii: 
\begin{equation}\label{eq:FX}
    \mathrm{VCF}_X=\dfrac{\Sigma Q_\mathrm{\hi}ZV_X/V_{\hii}}{\Sigma Q_\mathrm{\hi}Z}\,.
\end{equation}
We assume that the oxygen in each \hii\ region is either entirely singly-ionized or doubly-ionized so that:
\begin{equation}\label{eq:VCF_sum}
\mathrm{VCF}_\mathrm{\oii}+\mathrm{VCF}_\mathrm{\oiii}=1\,,
\end{equation}
which is reasonably accurate given the similar ionization thresholds for \oi\ and \hi.

Physically, these fractions are expected to depend on the shape of the photo-ionizing radiation spectrum: if most of the \hii\ regions in a galaxy see a hard ionizing spectrum, then $\mathrm{VCF}_\oiii$ will be close to unity, while 
galaxies with soft ionizing spectra will have $\mathrm{VCF}_\oii$ near one. The R23
diagnostic expression is relatively complicated, but is less sensitive to the VCFs since it varies with the sum of $\mathrm{VCF}_\oii$ and $\mathrm{VCF}_\oiii$.
The O3O2 and N2O2 diagnostics are metallicity independent to the first order\footnote{There is a ``higer-order'' dependence on metallicity via the gas temperature: the temperature is correlated with the metallicity, and so the temperature dependencies among O3O2 and N2O2 lead to indirect metallicity dependencies.}, but can help in determining the VCFs. For example, O3O2 varies with $\mathrm{VCF}_\oiii/\mathrm{VCF}_\oii$ which can then be combined with R2 or R3 to extract metallicities. As discussed further below, N2O2 can help break degeneracies when multiple metallicity solutions are allowed from the other strong line diagnostics.  

\section{Direct temperature measurements}\label{sec:directTe}

As alluded to in the Introduction, our models for R2, R3, R23, O3O2, and N2O2 rely on TZRs. More specifically, we assume a tight correlation between the effective \hii\ region metallicity, $Z_\mathrm{Q}$, of Eq.~(\ref{eq:diagnostic_T}) and the gas temperatures in the \oiii, \oii, \nii, and \hii\ regions in a given galaxy. We calibrate these TZRs using samples with direct $T_e$ temperature determinations, including those from
local galaxies \citep{2012A&A...546A.122I,2017A&A...599A..65G,2018MNRAS.473.1956I,2020A&A...634A.107Y}, galaxies at cosmic noon \citep{2018MNRAS.481.3520P,2023arXiv230603120L,2024ApJ...962...24S}, and some sources in the EoR \citep{2023arXiv230603120L,2023arXiv230112825N,2024ApJ...962...24S,2023MNRAS.518..425C}. The direct $T_e$ metallicity determinations from the literature will also be used as the ``ground truth'' to test our strong line diagnostic inferences. The local galaxy samples used in this work contain 216 systems ranging from individual \hii\ regions to the integrated emission from entire galaxies at $z<0.25$. The metallicities among this sample vary from $0.01Z_\odot$ to $1Z_\odot$. Among those, the 143 systems compiled by \cite{2020A&A...634A.107Y} include [\oiii] and [\oii] auroral line luminosity measurements for each source and so this allows clean determinations of the \oiii\ and \oii\ region gas temperatures for each line emitter. The [\oii] auroral line remains undetected in many low-metallicity local galaxies, and most of the high redshift galaxy samples. Hence for these samples we rely on spectral synthesis models to determine the \oii\ region temperatures from the more direct \oiii\ region temperature inferences (e.g. \citealt{2006A&A...448..955I}). Therefore, metallicity measurements without the [\oii] auroral line detection are only ``semi-direct''. The cosmic noon $(1<z<3)$ and EoR ($z>6$) samples considered in this work contain 15 and 20 line emitters, respectively. Both samples cover metallicities ranging approximately from $0.02Z_\odot$ to $0.5Z_\odot$. Note that in calibrating the TZRs, we assume the temperature and metallicity estimates reported in the literature, rather than re-deriving them.

\subsection{\oiii\ and \oii\ fractional volumes}\label{sec:ICFs}

\begin{figure}
    \centering
    \includegraphics[width=0.49\textwidth]{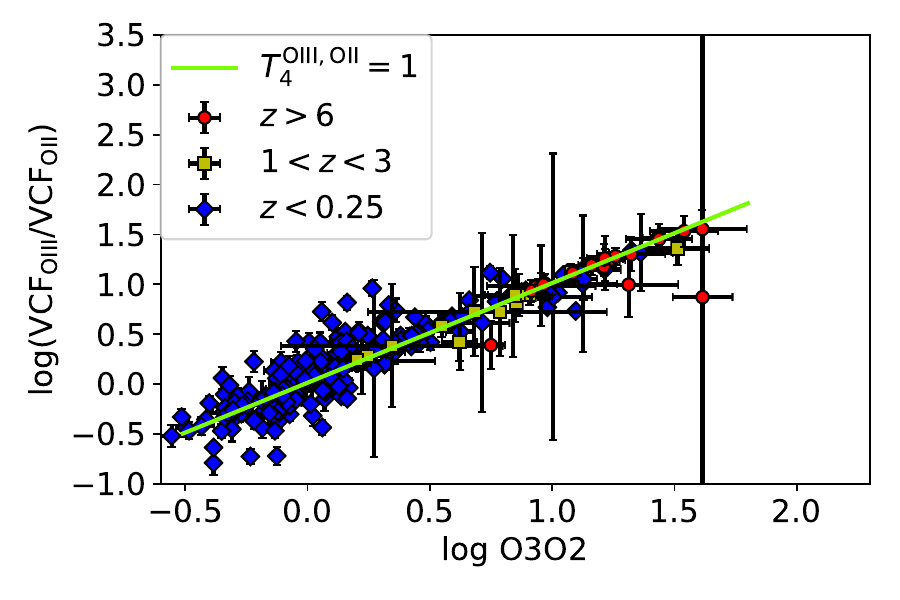}
    \includegraphics[width=0.49\textwidth]{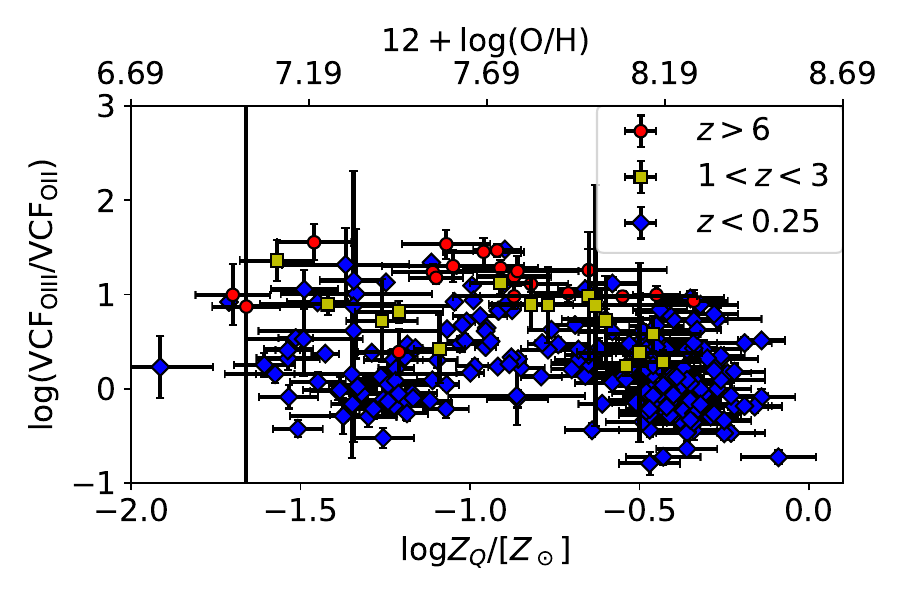}
    \caption{\oiii\ and \oii\ region VCFs versus O3O2 (top) and gas-phase metallicity (as defined in Eq.~\ref{eq:ZQ}, bottom). Red circles are high-$z$ measurements from \protect\cite{2023arXiv230603120L,2023arXiv230112825N,2024ApJ...962...24S,2023MNRAS.518..425C}. Yellow cubes are measurements at cosmic noon \protect\citep{2018MNRAS.481.3520P,2023arXiv230603120L,2024ApJ...962...24S}. Blue diamonds are local measurements from \protect\cite{2012A&A...546A.122I,2017A&A...599A..65G,2018MNRAS.473.1956I,2020A&A...634A.107Y}. 
    The green solid line in the top panel shows the expected relationship between O3O2 and $\mathrm{VCF}_\oiii$, $\mathrm{VCF}_\oii$ from Eq.~(\ref{eq:diagnostic_T}). The line assumes $T_4^{\rm{\oiii}} = T_4^{\rm{\oii}}=1$ but the results here are insensitive to the temperatures assumed (see text).
    The larger O3O2 values in higher redshift galaxies imply that a greater fraction of their volumes are in \oiii\ regions than in more nearby galaxies. This suggests that high-$z$ \hii\ regions are sourced by
    harder and possibly more intense ionizing spectra and/or that their \hii\ regions are denser (relative to more local galaxies). 
    }\label{fig:F32}
\end{figure}\par

As introduced in Section~\ref{sec:diagnostics}, Eq.~(\ref{eq:diagnostic_T}) allows us to determine $\mathrm{VCF}_\oiii/\mathrm{VCF}_\oii$ given measurements of O3O2, models or empirical fits for $T_4^\oiii$, $T_4^\oii$, and the assumption that $\mathrm{VCF}_\oiii+\mathrm{VCF}_\oii=1$. The quantity $\mathrm{VCF}_\oiii/\mathrm{VCF}_\oii$ in turn gives a handle on the hardness of the ionizing radiation spectrum in these galaxies. The top panel of Figure~\ref{fig:F32} shows the resulting derived values of $\mathrm{VCF}_\oiii/\mathrm{VCF}_\oii$, as a function of the O3O2 measurements, for all current direct and semi-direct $T_e$ estimates in the literature. The local, cosmic noon, and EoR measurements are shown as red circles, yellow cubes, and blue diamond points, respectively.
Since O3O2 is insensitive to temperature under the \oii\ temperature model assumed \citep{2006A&A...448..955I}, the trend
of O3O2 with the VCFs can be compactly described by:
\begin{equation}\label{eq:ICF_O3O2}
    \log(\mathrm{VCF}_\oiii/\mathrm{VCF}_\oii)=\log\mathrm{O3O2}+0.0128\,,
\end{equation}
as illustrated by the green line in Figure~\ref{fig:F32}. This simplified description captures the trends because O3O2 is insensitive
to temperature provided $T_4^\oiii \lesssim 2.5$, as 
is the case for all but three EoR, one cosmic noon source, and two local galaxies out of 252 total galaxies in the sample.

The figure illustrates that high-$z$ galaxies
have larger O3O2 galaxies and that a larger fraction of the oxygen in their \hii\ regions is doubly-ionized (rather than only singly-ionized).
This likely implies that the high redshift \hii\ regions are exposed to harder radiation spectra than in nearby galaxies. If the high-$z$ \hii\ regions have larger gas densities and/or feel more intense radiation fields, this could also play a role in explaining the observed trend (e.g. see Figure~1 from \defcitealias{2023MNRAS.tmp.2479Y}{Y23}\citetalias{2023MNRAS.tmp.2479Y}).  

The bottom panel of Figure~\ref{fig:F32} further considers the trend of VCFs with gas-phase metallicity (as inferred from direct $T_e$ measurements in the literature). Although the scatter is large, especially in the estimates from nearby galaxies, the higher redshift galaxies tend to have larger values of $\mathrm{VCF}_\oiii/\mathrm{VCF}_\oii$. This shows that the higher values of the \oiii\ to \oii\ fraction with increasing redshift may also occur at fixed gas-phase metallicity. This might
reflect increasing binarity among
high redshift stellar populations, which could lead to a harder radiation spectrum at fixed metallicity \citep{2020Galax...8....6S,2022ApJ...935..119S}. Another contributing factor may be a super-solar oxygen to iron abundance ratio at high redshift 
\citep{2016ApJ...826..159S}: in this context, this would imply that the effective stellar metallicity is lower than the gas-phase metallicity and this may lead to an increasingly hard and intense radiation field towards high-$z$. Finally, gas density evolution may be involved: for instance \citet{2014AAS...22322703M} and \citet{2016ApJ...816...23S}
argue that the typical gas densities in \hii\ regions among $z \sim 2$ galaxies are about an order of magnitude larger than in local galaxies with similar specific star-formation rates. 
\par

\subsection{\oii\ and \oiii\ region temperatures}\label{sec:Tgas}

\begin{figure*}
    \centering
    \includegraphics[width=0.49\textwidth]{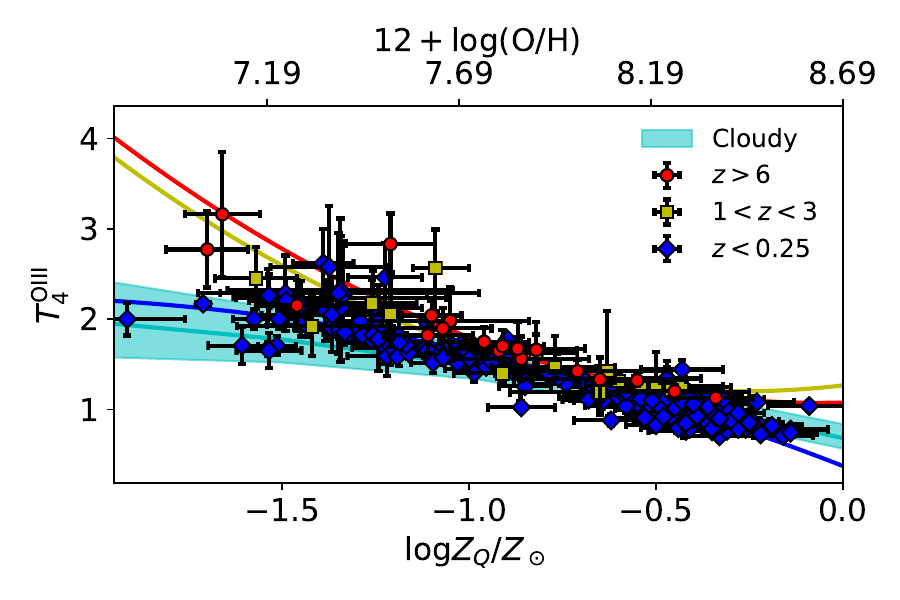}
    \includegraphics[width=0.49\textwidth]{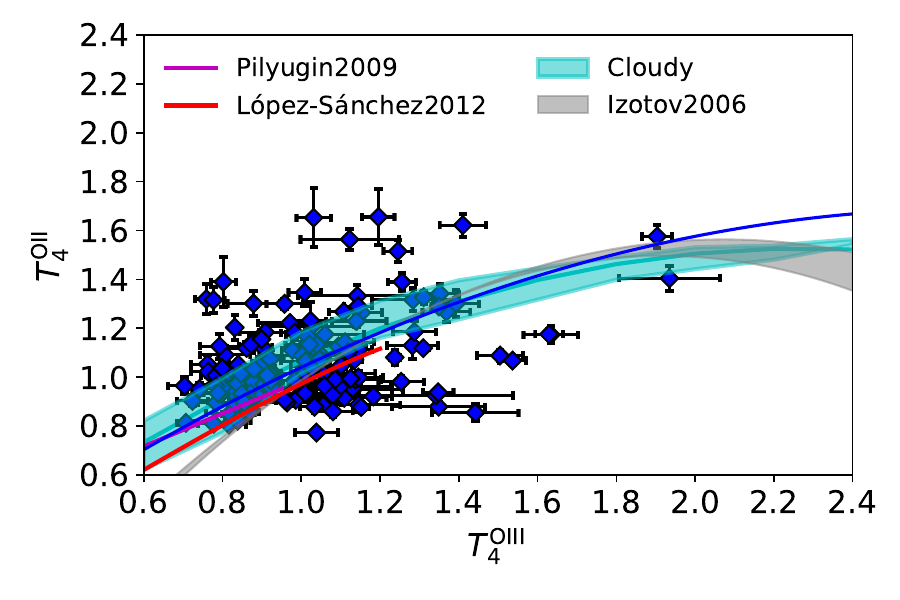}
    \caption{Theoretical and empirical correlations between the galaxy-averaged
    \oiii\ temperatures, \oii\ temperatures, and gas-phase metallicities. The empirical
    measurements come entirely from the direct $T_e$ method. The blue, yellow, and red data points are from the nearby, cosmic noon, and EoR galaxy samples, respectively. In each panel, the cyan band shows the theoretical expectations from \textsc{Cloudy} thermal equilibrium models (see text for details). 
    \textit{Left:} The \oiii\ temperature versus metallicity relation (TZR). The blue, yellow, and red curves show polynomial fits to the nearby, cosmic noon, and EoR galaxy samples, respectively. \textit{Right}: The \oii\ versus \oiii\ gas temperature correlation. The grey band shows the modeling results from \protect\cite{2006A&A...448..955I}, while the magenta and red curves show fits from  \protect\citet{2009MNRAS.398..485P} and \citet{2012MNRAS.426.2630L}, respectively, and the blue curve is a polynomial fit.}\label{fig:TZz}
\end{figure*}\par

Although the [\oiii] $\lambda 4364$~\AA\ auororal line is faint, \textit{JWST} has nevertheless detected it in tens of $z > 6$ galaxies. These measurements have allowed direct $T_e$ estimates of \oiii\  region gas temperatures and metallicities across a wide range of redshifts. Plots of these measurements are given in the left panel of Figure~\ref{fig:TZz} along with results from local and cosmic noon galaxies. The nearby, cosmic noon, and EoR estimates and associated error bars are shown, respectively, as blue diamonds, yellow squares, and red circles. The best fit TZRs for the local, cosmic noon, and EoR samples, derived from orthogonal distance regression, are shown as blue, yellow, and red curves. Specifically, these fits have the following functional forms:
\begin{equation}\label{eq:T4OIII_local}
    T_4^\oiii=-0.347\log Z_Q^2 -1.61\log Z_Q+  0.375\,,\ \mathrm{at}\ z<0.25\,,
\end{equation}
\begin{equation}\label{eq:T4OIII_cosmicNoon}
    T_4^\oiii=0.914\log Z_Q^2 +0.487\log Z_Q+  1.27\,,\ \mathrm{at}\ 1<z<3\,,
\end{equation}
\begin{equation}\label{eq:T4OIII_EOR}
    T_4^\oiii=0.824\log Z_Q^2 +0.101\log Z_Q+  1.08\,,\ \mathrm{at}\ z>6.
\end{equation}
Here the galaxy-wide gas-phase metallicity, $Z_Q$, is expressed in units of Solar metallicity. In order to better understand these results, we compute volume-averaged \oiii\ region temperatures using \textsc{Cloudy} models spanning a range of possible ISM parameters. The \textsc{Cloudy} calculations assume thermal equilibrium and adopt a spherical geometry. We consider models with $47\leq\log(Q_\mathrm{\hi}/[\mathrm{s^{-1}}])\leq51$, $2\leq\log n_\mathrm{H}/[\mathrm{cm^{-3}}]\leq4$,
and instantaneous stellar radiation spectra with an age of age 1--30 Myr \citep{2009ApJ...699..486C,2010ApJ...712..833C}. 
Here we assume that the gas-phase and effective stellar metallicity are identical. The above parameter ranges were chosen because \defcitealias{2023MNRAS.tmp.2479Y}{Y23}\citetalias{2023MNRAS.tmp.2479Y} found that these parameter values account for most of the \oiii, \oii, and Balmer line emission in their model star-forming galaxies. Here we define $T_4^X$ in each \textsc{Cloudy} model as the volume-averaged gas temperature for ion $X$. Interestingly, the simple thermal equilibrium \textsc{Cloudy} models mostly agree with the direct $T_e$ estimates, although the data provide a hint that the lowest metallicity galaxies (at log($Z_{\mathrm Q}/Z_\odot) \lesssim -1.25)$ may have slightly warmer \oiii\ regions than expected from the models. Also, current high-$z$ direct $T_e$ measurements (yellow and red scatters) lie along the upper envelope of the thermal equilibrium TZRs for $Z\gtrsim0.1Z_\odot$. Overall, the results suggest that the \hii\ regions in these galaxies may generally be close to thermal equilibrium, especially in the nearby samples.

The mild TZR redshift evolution could again reflect a harder ionizing spectrum at fixed metallicity in the high-$z$ sources, which would lead to more photo-heating and hotter \hii\ regions. Alternatively, there could be a selection effect if galaxies with hot \oiii\ regions and bright [\oiii] auroral lines are preferentially selected in the high-$z$ samples. Finally, the result might be related to the possibility that star-formation at high redshift 
is burstier, which could lead to departures from thermal equilibrium \citep{2023MNRAS.525.3254S,2023ApJ...955L..35S}. 

Unfortunately, [\oii] auroral lines (with restframe wavelengths near $7330$~\AA) are even fainter and hence generally only detected in nearby galaxy samples. 
Consequently, the right panel of Figure~\ref{fig:TZz} shows direct $T_e$ estimates
for $T_4^\mathrm{\oii}$ as a function of $T_4^\mathrm{\oiii}$ but only among 143 nearby galaxies and \hii\ regions with [\oii] auroral line detections \citep{2020A&A...634A.107Y}.
The right panel of Figure~\ref{fig:TZz}
compares \textsc{Cloudy} thermal equilibrium calculations of the \oii\ and \oiii\ region temperatures (cyan band) with other models
in the literature \citep{2006A&A...448..955I,2009MNRAS.398..485P,2012MNRAS.426.2630L}, and local direct $T_e$ measurements (blue diamonds). The best fit from an orthogonal distance regression follows: 
\begin{equation}\label{eq:T4OII}
    T_4^\oii=-0.219(T_4^\oiii)^2+1.19T_4^\oiii+0.0666\,,
\end{equation}
as shown by the blue curve. The \textsc{Cloudy} thermal equilibrium models capture the general trend suggested by the direct $T_e$ measurements, but the scatter in the data is fairly large.\par
In this work we assume that the \hii\ region temperature matches the volume-average of the \oiii\ and \oii\ region temperatures:
\begin{equation}\label{eq:THII}
    T_4^\hii=T_4^\oiii \mathrm{VCF}_\oiii+T_4^\oii \mathrm{VCF}_\oii\,.
\end{equation}

\begin{figure}
    \centering
    \includegraphics[width=0.49\textwidth]{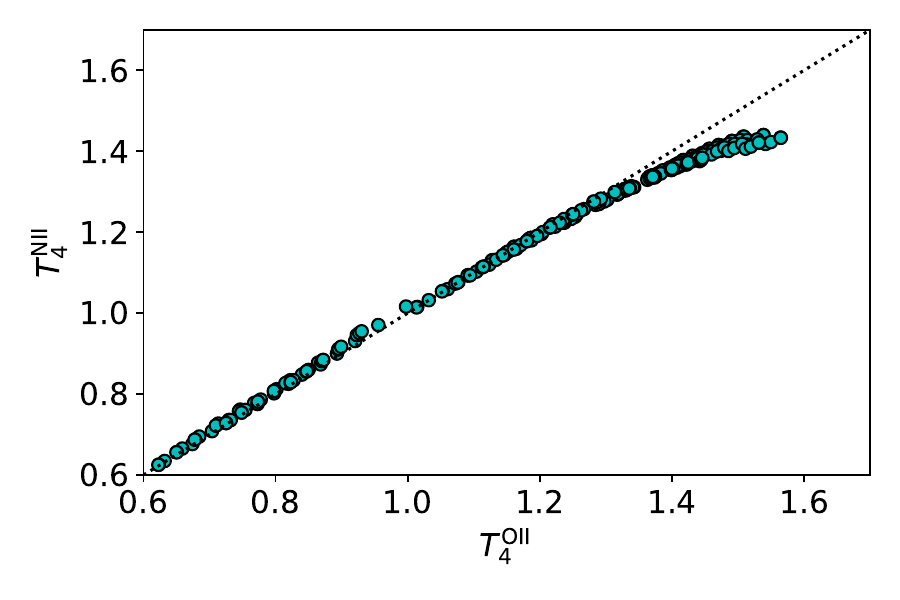}
    \includegraphics[width=0.49\textwidth]{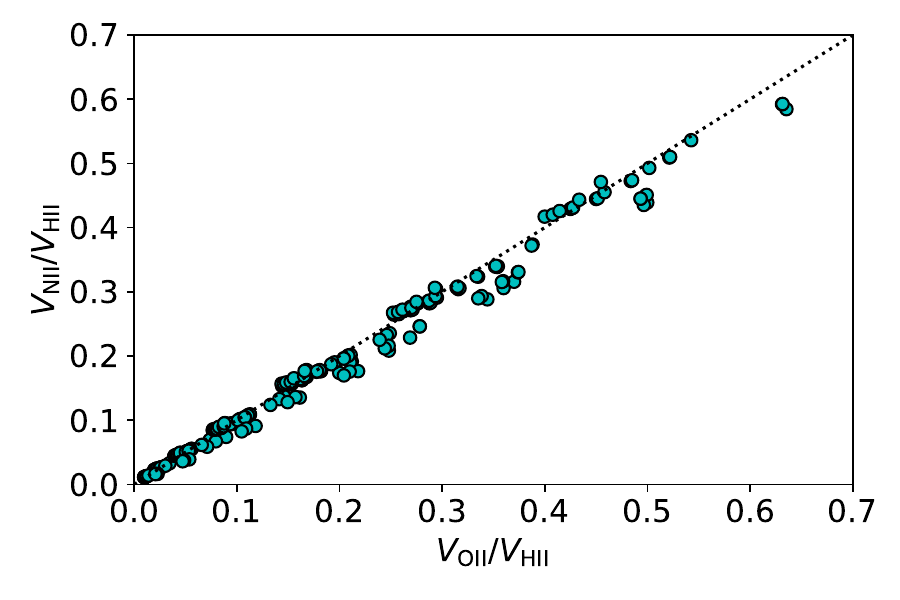}
    \caption{The connection between \nii\ and \oii\ regions determined from 300 \textsc{Cloudy} runs after varying the ISM parameters and incident stellar radiation spectra. \textit{Top:} A comparison between the volume-averaged temperatures of the \nii\ and \oii\ regions. 
    \textit{Bottom:} A comparison between the fraction of the volume of the \hii\ regions in each of \nii\ and \oii. The black dotted curve in each panel indicates equal temperatures/equal fractional volumes. The \nii\ and \oii\ regions are expected to have similar gas temperatures and fractional volumes. 
    }\label{fig:Cloudy_NII}
\end{figure}\par

Finally, since we also consider N2O2 it is important to compare the temperature and fractional volumes of the \nii\ and \oii\ regions. Our \textsc{Cloudy} runs suggest that $T_4^\nii=T_4^\oii$, $F_\nii=\mathrm{VCF}_\oii$ are good approximations across the parameter space of interest, as illustrated in Figure~\ref{fig:Cloudy_NII}. Here $F_X$ in each \textsc{Cloudy} model is simply the fractional volume occupied by ion $X$ within the \hii\ region. Note that $T_4^\nii\approx T_4^\oii$ is also found empirically in local direct $T_e$ measurements (e.g. \citealt{2017MNRAS.465.1384C}). In this work we assume $T_4^\nii=T_4^\oii$, $F_\nii=\mathrm{VCF}_\oii$ so that N2O2 in Eq.~(\ref{eq:diagnostic_T}) can be simplified as: 
\begin{equation}\label{eq:N2O2}
    \log(\mathrm{N2O2})=\log\left[\dfrac{0.0738k_{03}^\mathrm{\nii}(T_4^\mathrm{\oii})}{k_{01}^\mathrm{\oii}(T_4^\mathrm{\oii})+k_{02}^\mathrm{\oii}(T_4^\mathrm{\oii})}\right]\,.
\end{equation}
In this approximation, N2O2 is only a function of the temperature of each \oii\ region.

\section{Constraining metallicities with strong line diagnostics}\label{sec:Obs}

\begin{figure*}
    \centering
    \includegraphics[width=0.49\textwidth]{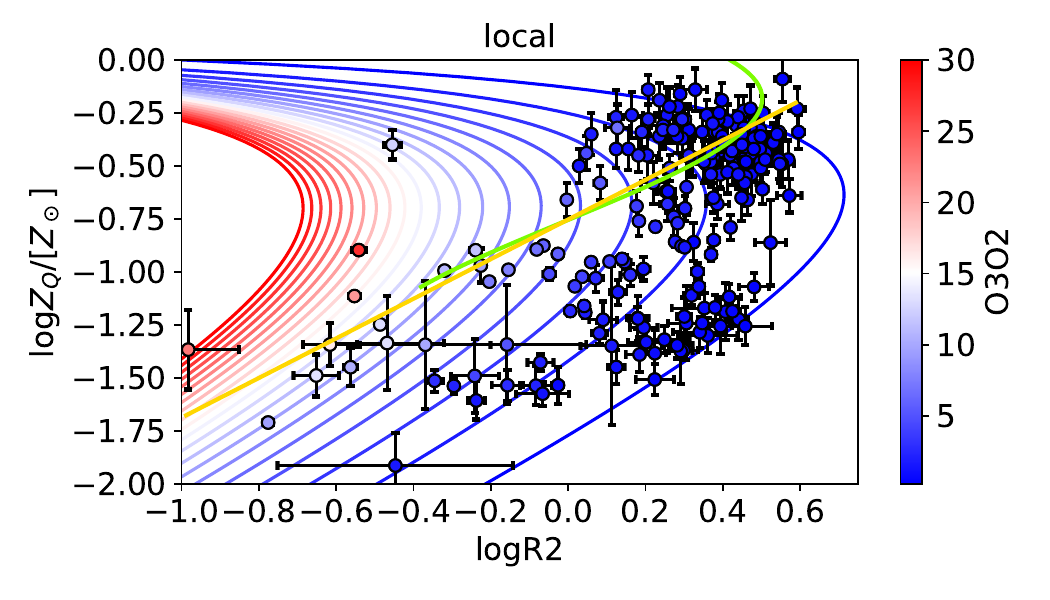}
    \includegraphics[width=0.49\textwidth]{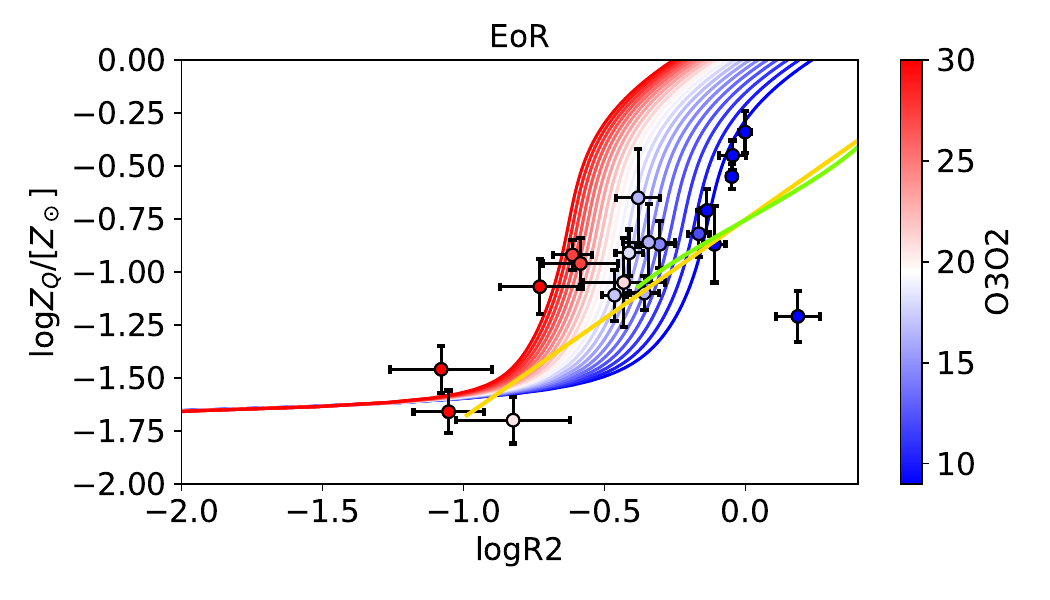}\\
    \includegraphics[width=0.49\textwidth]{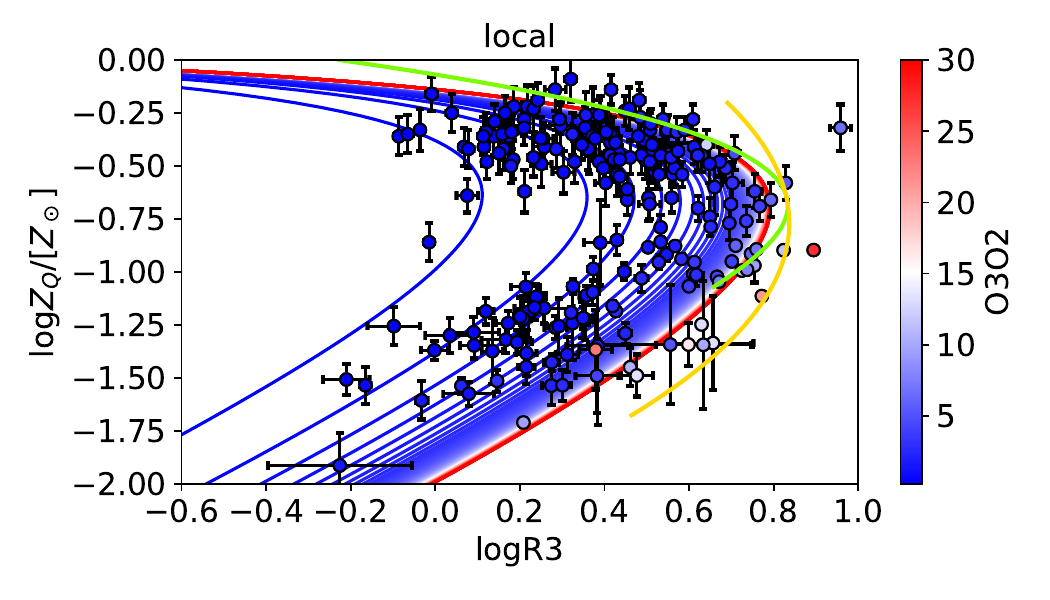}
    \includegraphics[width=0.49\textwidth]{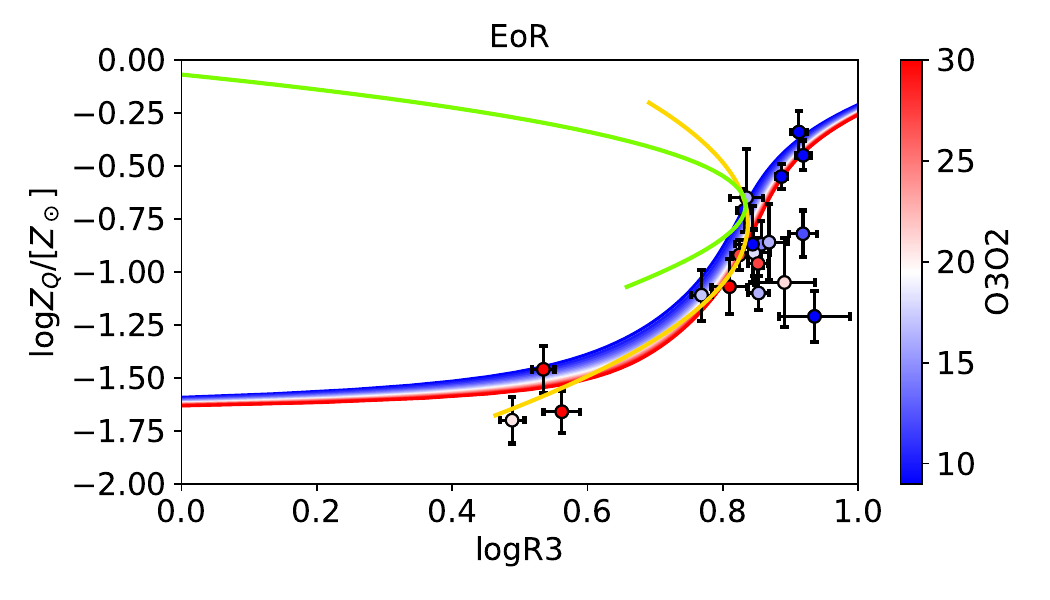}\\
    \includegraphics[width=0.49\textwidth]{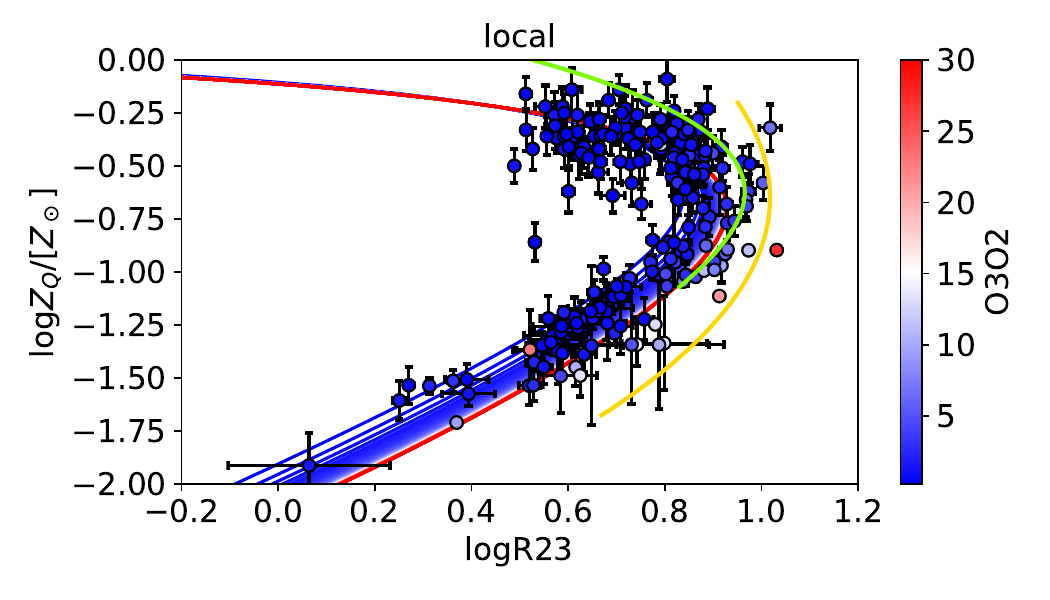}
    \includegraphics[width=0.49\textwidth]{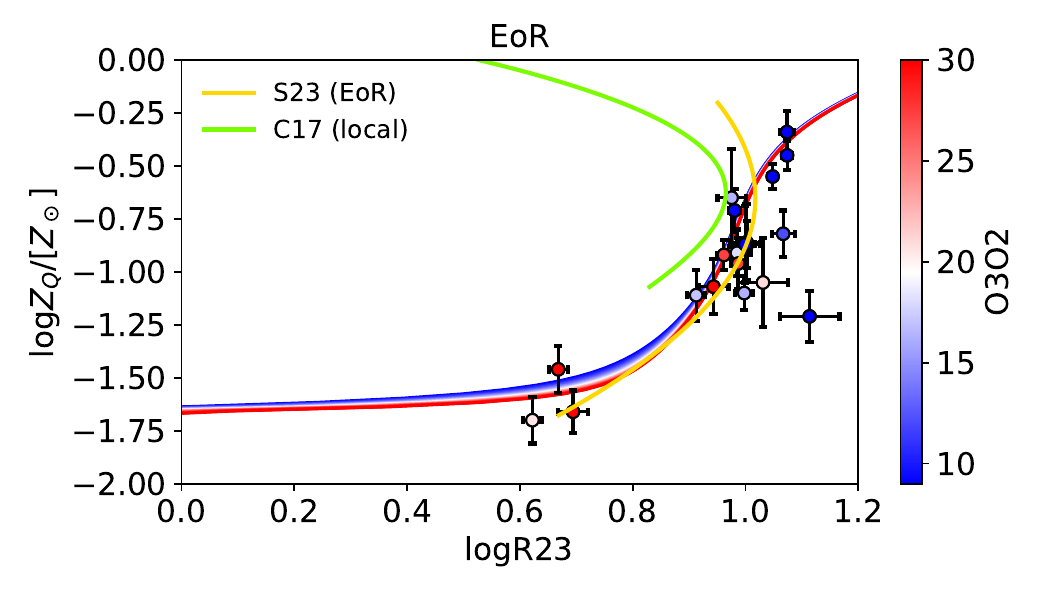}\\
    \caption{Metallicity versus R2 (top row), R3 (middle row), and R23 (bottom row) as a function of O3O2 (color bar). 
    The left column adopts \oiii\ and \oii\ TZRs based on local measurements
    (Eqs.~\ref{eq:T4OIII_local} and \ref{eq:T4OII}). The right column assumes
    the high-$z$ \oiii\ TZR of Eq.~(\ref{eq:T4OIII_EOR}) calibrated from direct $T_e$ measurements, while the \oii\ temperatures are determined using
    the fitting formula from \protect\cite{2006A&A...448..955I}. The points
    (color-coded by O3O2) in the left (right) column are from local (EoR) direct $T_e$ measurements. The green and gold curves show empirical strong line diagnostic calibrations from local \citep{2017MNRAS.465.1384C}
    and EoR \citep{2024ApJ...962...24S} galaxy samples, respectively. Much of the scatter among the strong line diagnostic ratio measurements is explained by variations in O3O2 across different galaxies in the samples. The changes from the local to EoR galaxy samples are largely driven by redshift evolution in O3O2 and in the gas temperatures. 
    }\label{fig:R}
\end{figure*}\par

Under the assumptions of our model, Eq.~\ref{eq:diagnostic_T} can be used
to determine R2, R3, R23, O3O2, and N2O2 given the gas-phase metallicity, $Z_\mathrm{Q}$, the VCF, $\mathrm{VCF}_\oiii/\mathrm{VCF}_\oii$, and the TZRs. From measurements of R2, R3, R23, O3O2, and N2O2 we can turn this around and extract constraints on $Z_\mathrm{Q}$ and $\mathrm{VCF}_\oii/\mathrm{VCF}_\oii$. For example, consider the case of applying the R2 and O3O2 diagnostics alone. Eq~\ref{eq:diagnostic_T} shows that R2 and O3O2 depend on 6 total parameters in our model: $T_\mathrm{OIII}$, $T_\mathrm{OII}$, $T_\mathrm{HII}$, $\mathrm{VCF}_\mathrm{OIII}$, $\mathrm{VCF}_\mathrm{OII}$, and $Z_Q$. In our model the OII and OIII temperatures are fully specified by the TZRs (i.e. they are determined by $Z_Q$). The HII region temperature follows from the volume fractions of OII and OIII and their 
temperatures (Eq~\ref{eq:THII}). Eq~\ref{eq:VCF_sum} eliminates another degree-of-freedom. This leaves two independent parameters which can be determined from the two observables, R2 and O3O2.\par 
The accuracy of this technique can be cross-checked using samples with both strong line diagnostic measurements and direct $T_e$ determinations of $Z_\mathrm{Q}$.  
Here we show model predictions for $Z_\mathrm{Q}(\mathrm{R2},\mathrm{O3O2})$, $Z_\mathrm{Q}(\mathrm{R3},\mathrm{O3O2})$, and $Z_\mathrm{Q}(\mathrm{R23},\mathrm{O3O2})$ for each of the two temperature versus metallicity models of
Section~\ref{sec:Tgas}, which are calibrated from current local and high-$z$ galaxy measurements, respectively. We then extract strong line diagnostic metallicity estimates using our model from R2, R3, R23, or O3O2 measurements and compare them to direct $T_e$ determinations and other results in the literature. Our results can be further refined in the future with improved
empirical determinations and/or theoretical models of the TZRs. 

For the nearby galaxy samples, we assume that the \oiii\ and \oii\ region gas temperatures are determined by the effective gas-phase metallicity, $Z_\mathrm{Q}$, according to Eqs.~(\ref{eq:T4OIII_local}) and (\ref{eq:T4OII}).
The corresponding model predictions for the metallicity as a function of the strong line diagnostic ratios, $Z_\mathrm{Q}(\mathrm{R2})$, $Z_\mathrm{Q}(\mathrm{R3})$, and $Z_\mathrm{Q}(\mathrm{R23})$ , are shown in the left column of Figure~\ref{fig:R}, and color-coded by the value of O3O2. 
The corresponding measurements for the nearby galaxies in the sample are shown by the points with error bars in the left column panels. 
Considering first the R2 models, note that at fixed R2 and O3O2, two metallicity solutions are possible. This occurs because R2 first increases with metallicity, but then subsequently declines as the more metal rich gas is cooler and this leads to less collisional excitation and [\oii] emission. 
(The decline in [\oii] luminosity with gas temperature is stronger than the more mild drop in the H$\beta$ luminosity -- which enters in the denominator of R2 -- with temperature). For similar reasons, R3 and R23 also show a double-valued dependence on metallicity. As O3O2 gets larger, the 
fractional sizes of the \oiii\ (\oii) regions increase (decrease), resulting in greater values of R3 (and smaller R2) at a fixed $Z_\mathrm{Q}$. The R23 diagnostic, on the other hand, is less sensitive to $\mathrm{VCF}_\oiii$ and $\mathrm{VCF}_\oii$ since it depends on a sum of these quantities (see Eq.~\ref{eq:diagnostic_T}). We contrast our model curves with empirically-based fits to local measurements from the literature \citep{2017MNRAS.465.1384C}, as shown in the green curves. These fits ignore the O3O2 dependence of R2, R3, and R23, although this can account for much of the scatter in the measurements. For contrast, the gold curve shows empirical fits to current EoR samples \citep{2024ApJ...962...24S}, which illustrates the level of redshift evolution found in the current empirical calibrations. 

The right panel of Figure~\ref{fig:R} shows the same strong line diagnostic plots, except here for $z > 6$ galaxies. In this case we assume the TZR from
Eq.~(\ref{eq:T4OIII_EOR}) for the \oiii\ temperature and the relationship from \cite{2006A&A...448..955I} to determine the \oii\ region temperature from the \oiii\ one. The resulting $Z_\mathrm{Q}(\mathrm{R2})$, $Z_\mathrm{Q}(\mathrm{R3})$, and $Z_\mathrm{Q}(\mathrm{R23})$ models (right panel) no longer show the double-valued behavior evident at lower redshifts. Instead, in this case, R2, R3, and R23 increase monotonically with metallicity. This is the case because under the high-$z$ TZR model, the increase in line ratio with metallicity dominates over the effect from the decline in temperature with growing metallicity. \par

We assume temperature model Eq.~(\ref{eq:T4OIII_cosmicNoon}) and \cite{2006A&A...448..955I} for galaxies near cosmic noon. Since this temperature model is very similar to that of EoR galaxies, the cosmic noon strong line diagnostics predicted by our method is similar to relations presented in Figure~\ref{fig:R} right column. For brevity, we do not show plots for cosmic noon line ratio modeling results.\par
Although our model curves mostly pass through the strong line ratio measurement data, there are a few outliers that are not well-explained by our model. For example, in the EoR sample the galaxies at large R3 and R23, especially those at low metallicity, are unexpected in our model. This is caused by the fact that our model assumes a one-to-one relationship between effective gas phase metallicity, $Z_\mathrm{Q}$, and the \oiii\ and \oii\ region temperatures, and so neglects scatter in these correlations. In future work, it may be interesting to refine our calculations to incorporate scatter in the TZRs. Another factor is that in this work we have assumed $T_4^\oii(T_4^\oiii)$ relations Eq~\ref{eq:T4OII} and \cite{2006A&A...448..955I} for local and high-z galaxies, while different observational groups may have assumed slightly different $T_4^\oii(T_4^\oiii)$ relations (e.g. \citealt{1986MNRAS.223..811C,2009MNRAS.398..485P}) when the [\oii] auroral line is not detected.

\begin{figure*}
    \centering
    \includegraphics[width=1\textwidth]{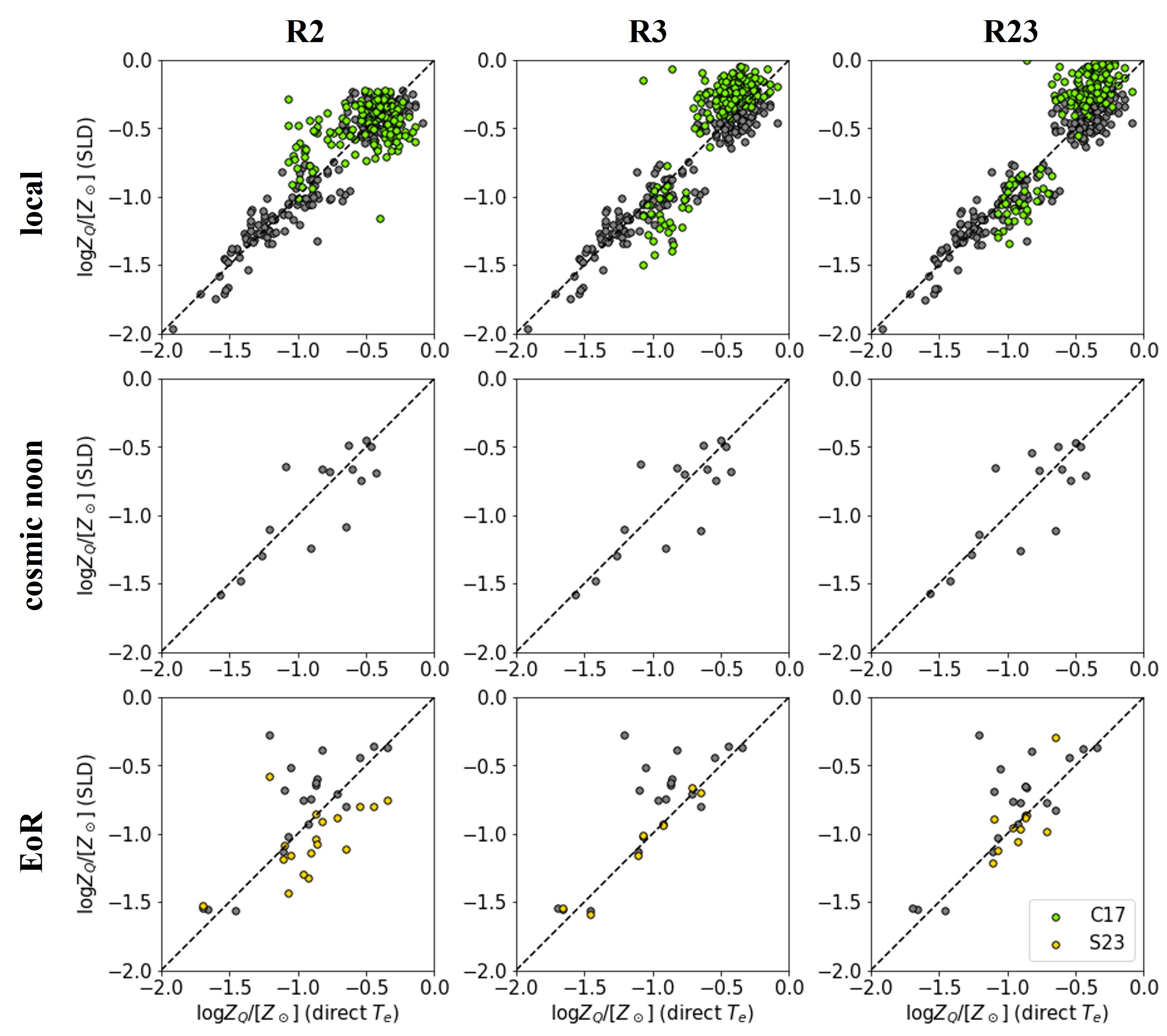}
    \caption{A comparison between the success of our strong line diagnostic models and empirical methods in the literature. The best-fit metallicities from our approach (grey) and empirical relations \protect\cite{2017MNRAS.465.1384C} (green), \protect\cite{2024ApJ...962...24S} (gold) are plotted against the best-fit direct $T_e$ metallicity determinations (for the same galaxies). Although the strong line metallicity estimates can be double-valued, we show only the estimate which lies closest to the corresponding direct $T_e$ metallicity measurements (see text). \textit{Top row:} The metallicity estimates from the low-$z$ galaxy sample. \textit{Middle row:} The same for the $1<z<3$ galaxy sample. \textit{Bottom row:} The same for the $z > 6$ galaxies.
    The left, middle, and right columns show determinations from R2, R3, and R23 (the metallicity is determined separately from each line ratio after combining with O3O2). The dotted line in each panel shows the one-to-one relation expected for perfect agreement between each strong line diagnostic metallicity and a direct $T_e$ determination. Overall, our method performs better than the empirical calibrations in the literature and allows strong line metallicity determinations for additional galaxies (since both methods sometimes fail to find an acceptable metallicity given the strong line measurements; see text). 
    }\label{fig:compare}
\end{figure*}\par

Figure~\ref{fig:compare} compares the performance of our new method (grey points) with direct $T_e$ measurements, as well as other, empirically-based, strong line diagnostic calibrations from \cite{2017MNRAS.465.1384C} for local galaxies (green, top row), and \cite{2024ApJ...962...24S} for EoR galaxies (gold points, bottom row).
We show separate metallicity determinations for each pairing of 
$\{\mathrm{R2,O3O2}\}$, $\{\mathrm{R3,O3O2}\}$, and $\{\mathrm{R23,O3O2}\}$. To the best of our knowledge, currently there is no strong line diagnostic model calibrated dedicatedly to cosmic noon galaxies. We therefore do not compare our cosmic noon model performance with other methods here.
As discussed previously, in some cases there can be two separate metallicity solutions for a given pair of strong line diagnostics, such as $\{\mathrm{R2,O3O2}\}$. In the case that there are two solutions, we show only the one closer to the direct $T_e$ metallicity estimate. In practice, N2O2 can help in breaking the degeneracy between the two branches of solutions: high N2O2 prefers lower gas temperature and the upper metallicity branch, while low N2O2 favors higher gas temperature and the lower metallicity solution. Alternatively, if stellar mass estimates are available, even a conservative prior on the mass--metallicity relation may favour one potential metallicity solution over the other. 
In cases where our model initially fails to find a solution, we allow
the \oiii\ and \oii\ region temperatures to increase by 2000 K. This is motivated
by the existence of scatter around the assumed TZRs, with empirical studies suggesting that 2000 K is a plausible level of intrinsic scatter (e.g. \cite{2009ApJ...700..309B,2013ApJ...777...96C,2023NatAs.tmp...87C}). The temperature adjustment then partly accounts for scatter around the assumed TZRs. 

\begin{table}
\centering
\begin{tabular}{|llll|}
\hline
\multicolumn{4}{|c|}{local}                                                                                        \\ \hline
\multicolumn{1}{|l|}{}                 & \multicolumn{1}{l|}{R2}        & \multicolumn{1}{l|}{R3}        & R23       \\ \hline
\multicolumn{1}{|l|}{$N_\mathrm{gal}$} & \multicolumn{1}{l|}{215/146}   & \multicolumn{1}{l|}{215/161}   & 216/151   \\ \hline
\multicolumn{1}{|l|}{$\sigma(\log Z_Q)$} & \multicolumn{1}{l|}{0.14/0.23} & \multicolumn{1}{l|}{0.14/0.24} & 0.14/0.24 \\ \hline
\multicolumn{4}{|c|}{cosmic noon}                                                                                       \\ \hline
\multicolumn{1}{|l|}{}                 & \multicolumn{1}{l|}{R2}        & \multicolumn{1}{l|}{R3}        & R23       \\ \hline
\multicolumn{1}{|l|}{$N_\mathrm{gal}$} & \multicolumn{1}{l|}{15/-}     & \multicolumn{1}{l|}{15/-}     & 15/-     \\ \hline
\multicolumn{1}{|l|}{$\sigma(\log Z_Q)$} & \multicolumn{1}{l|}{0.21/-} & \multicolumn{1}{l|}{0.22/-} & 0.22/- \\ \hline
\multicolumn{4}{|c|}{EoR}                                                                                       \\ \hline
\multicolumn{1}{|l|}{}                 & \multicolumn{1}{l|}{R2}        & \multicolumn{1}{l|}{R3}        & R23       \\ \hline
\multicolumn{1}{|l|}{$N_\mathrm{gal}$} & \multicolumn{1}{l|}{20/17}     & \multicolumn{1}{l|}{20/7}     & 20/11     \\ \hline
\multicolumn{1}{|l|}{$\sigma(\log Z_Q)$} & \multicolumn{1}{l|}{0.30/0.30} & \multicolumn{1}{l|}{0.30/0.075} & 0.30/0.16 \\ \hline
\end{tabular}
\caption{
A comparison between the performance of the
strong line diagnostic method introduced here and empirical calibrations from the literature. 
Here $N_\mathrm{gal}$ gives the number of galaxies for which the metallicity may be determined using strong line diagnostics. The standard deviation $\sigma(\log Z_Q)$ is defined in Eq.~(\ref{eq:chi}) and gives the average difference (in dex) between the strong line diagnostic and direct $T_e$ metallicity estimates. The first
number in each cell gives the result for our model while the second number is for \protect\cite{2017MNRAS.465.1384C} (low-$z$) or \protect\cite{2024ApJ...962...24S} (high-$z$). Currently there lacks strong line diagnostic model fitted dedicatedly to cosmic noon targets, and we do not compare our method with other works at $1<z<3$. Our model
performs better for the local galaxy samples. At high redshift, the method of \protect\cite{2024ApJ...962...24S} gives smaller fractional errors for R3 and R23, but our technique finds metallicities for a larger number of galaxies and performs better for R2. 
}
\label{tab:compare}
\end{table}

In order to compare the different strong line methods quantitatively, we compute the standard deviation between metallicity determined from strong line diagnostics and the direct $T_e$ estimate:
\begin{equation}\label{eq:chi}
    \sigma(\log Z_Q)=\sqrt{\dfrac{\Sigma(\log Z_{Q,\mathrm{SLD}}-\log Z_{Q,\mathrm{direct}})^2}{N_\mathrm{gal}}}\,.
\end{equation}
Here, $N_\mathrm{gal}$ is the number of galaxies
with at least one metallicity solution from the strong line ratio measurement. Table~\ref{tab:compare} gives the values of $\sigma(\log Z_Q)$ from our method and empirical fits in the literature for each diagnostic R2, R3, and R23 (when combined with O3O2). This comparison for the low-$z$ samples, with empirical strong line diagnostics from \cite{2017MNRAS.465.1384C}, and at high-$z$ using the results of \cite{2024ApJ...962...24S}. Due to the limited metallicity calibration ranges of the empirical models, we only apply  the \cite{2017MNRAS.465.1384C} approach to 163 local galaxies with $-1.1\leq\log Z_{Q,\mathrm{direct}}\leq0.16$, and apply the \cite{2024ApJ...962...24S} method to 20 EoR galaxies with $-1.7\leq\log Z_{Q,\mathrm{direct}}\leq-0.19$. In contrast, the method introduced in this work is applied to the entire set of 216 local galaxies, 15 cosmic noon galaxies, and 20 EoR galaxies introduced in Section~\ref{sec:directTe}. For the low redshift samples, our strong line diagnostic results are within $0.14$ dex of the direct $T_e$ metallicities, on average, for each of R2, R3, and R23. This compares favourably to the results from \cite{2017MNRAS.465.1384C}, which on average gives metallicities $\sim0.24$ differently from the direct $T_e$ measurements. In addition, our method finds metallicity solutions for a slightly larger number of galaxies at low redshift. At $z > 6$, the performance is generally not as good with our metallicities differing from the direct $T_e$ estimates by $\sim$0.30\,dex. In addition, the average performance of the empirical method of \cite{2024ApJ...962...24S} is generally better than our approach here. However, our calculations find metallicities for a larger number of galaxies, especially in the case of the R3 and R23 diagnostic. The $\sigma(\log Z_Q)$ given by our methods among the EoR galaxies with metallicity solutions in \cite{2024ApJ...962...24S} are 0.32, 0.082, and 0.19 dex for the R2, R3, and R23 diagnostics, respectively. In this sense, the performance of our method is comparable to \cite{2024ApJ...962...24S}. As more reionization-era direct $T_e$ measurements become available in the near future, our method can be refined by establishing more accurate TZRs and may ultimately outperform the empirically calibrated strong line diagnostics, as it currently does among low redshift samples. At $1<z<3$, our method is still able to find metallicity solutions for all galaxies. Metallicities constrained by our model is on average different from the direct $T_e$ measurements by $\sim$0.22\,dex.

\section{Discussion and Conclusions}\label{sec:summary}

The strong line diagnostics discussed in this work provide convenient metallicity estimates when auroral lines are undetectable and the gas
temperature remains undetermined. In this work we constructed models for common strong line diagnostic ratios from first principles. Our modeling clarifies how strong line diagnostic ratios depend on gas temperature, and gives a more precise definition of the ``metallicity'' constrained by strong line measurements (see Eq.~\ref{eq:ZQ}). We showed
how analytic models for [\oiii], [\oii], and H$\beta$ line emission can be combined with empirical fits describing correlations between gas temperature and metallicity to extract metallicities from strong line ratios including R2, R3, R23, and O3O2. Overall, as quantified in the previous section, our method performs comparably or better than empirically-motivated results in the literature. Furthermore, we believe our technique makes a more transparent set of assumptions. Here we briefly expand on the assumptions, strengths, and limitations of our new methodology. 

\subsection{Further cross-checks on the assumptions in this work}

First, we examine the assumption made here that the \hii\ region gas densities
are always much smaller than the critical densities of the emission lines considered. In the case of the
[\oiii] 4960\,\AA, 5008\,\AA, [\nii] 6584\,\AA, H$\alpha$, and H$\beta$ lines, the critical densities are larger than $\sim 10^5$ cm$^{-3}$.
Given that the typical \hii\ region gas densities range from a few to $\sim 10^3$ cm$^{-3}$ 
\citep{2014AAS...22322703M,2016ApJ...816...23S},
the low density limit is well-justified for these lines. However, the critical densities for the [\oii] 3727,30\,\AA\ doublet are smaller, $\sim$ a few $\times 10^3$ cm$^{-3}$ and so this approximation deserves closer scrutiny (as discussed further below). Second, we assume that all of the oxygen within an \hii\ region sourced by stellar radiation is either singly or doubly-ionized. That is, $\mathrm{VCF}_\oiii+\mathrm{VCF}_\oii=1$, and we ignore the possible presence of neutral oxygen and more highly-ionized oxygen. Finally, we neglect temperature variations across the \hii\ regions in each galaxy and any scatter around the average temperature-metallicity correlation, which could cause metallicity underestimation \citep{2023MNRAS.522L..89C}. These second and third assumptions -- i.e., that the oxygen in \hii\ regions is entirely singly or doubly-ionized and that temperature fluctuations are negligible -- are widely adopted. The assumption regarding the ionization state of oxygen is supported by the theoretical models of \cite{2020MNRAS.499.3417Y} and \defcitealias{2023MNRAS.tmp.2479Y}{Y23}\citetalias{2023MNRAS.tmp.2479Y}. There is at least some observational support for neglecting temperature fluctuations \citep{2023NatAs.tmp...87C}. Under these three assumptions, our expressions for the strong line ratios in Eq.~(\ref{eq:diagnostic_T}) follow, and one can solve for the metallicity given the gas temperature.

\begin{figure}
    \centering
    \includegraphics[width=0.49\textwidth]{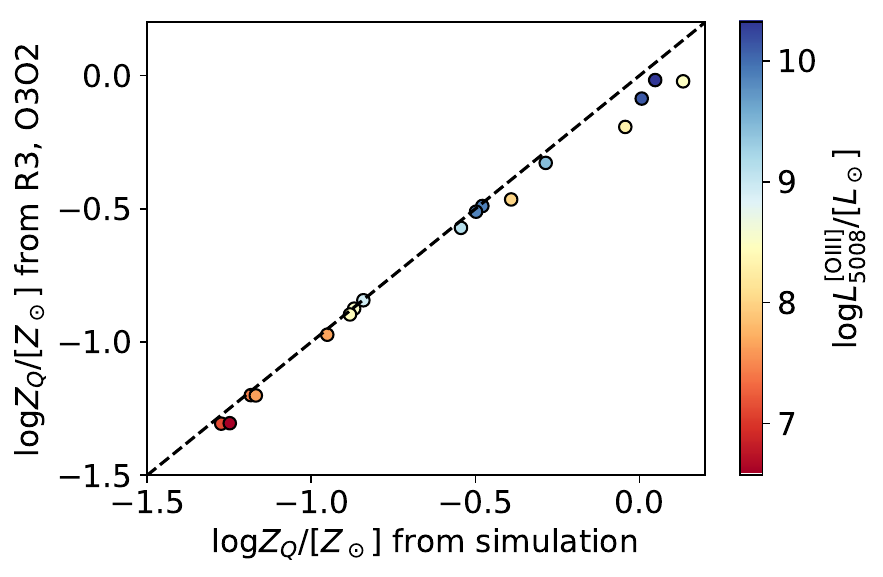}
    \caption{
    A test of the low gas density approximation
    using mock emission lines from galaxies in the FIRE simulations. The y-axis shows the metallicity ($Z_Q$) inferred from 17 simulated [\oii]-emitting FIRE galaxies and the R3 and O3O2 diagnostics in the low density limit, compared to the true metallicity from the simulations. Each simulated galaxy is color-coded by its [\oiii]\,5008\,\AA\ line luminosity.
    The black dashed line shows the case of perfect agreement. In all cases, the low density approximation is good to better than 0.16 dex (and in most of the simulated galaxies the agreement is better than this).
    }\label{fig:logZQ_FIRE}
\end{figure}

Figure~\ref{fig:logZQ_FIRE} provides a quantitative test of the low gas density approximation. Specifically, we use the full [\oii] line emission models, post-processed onto 17 of the FIRE high-$z$ galaxies with $L^{[\oii]}_{3727,30}\gtrsim10^6 L_\odot$ in \defcitealias{2023MNRAS.tmp.2479Y}{Y23}\citetalias{2023MNRAS.tmp.2479Y}. Since
the FIRE modeling efforts include the effect of
collisional de-excitations, i.e. they do not assume the low density limit, these studies can help us cross-check this approximation. 
The figure reveals that the R3 and O3O2-derived metallicities (assuming the low density approximation) are close but slightly lower than the true metallicities ($Z_Q$). Ignoring collisional de-excitations in the strong line diagnostics causes one to slightly overestimate the [\oii] line luminosity, and this in turn leads to a small underestimation of the metallicity. However, this effect is always smaller than 0.16 dex in the FIRE models. This is smaller than current strong line diagnostic metallicity uncertainties (see Table~\ref{tab:compare}) and so we believe the low density approximation is adequate for present purposes. Note that the low density approximation does not always apply to individual HII regions, where the gas density can be as high as $10^5$ cm$^{-3}$ (e.g. \citealt{2023MNRAS.523.2952M}), but we believe it should be accurate when applied across the ensemble of emitting HII regions in an entire galaxy.

Another potential concern with our method relates
to the empirical calibrations adopted for the
gas temperatures. Specifically, in Section~\ref{sec:Tgas} we fit direct $T_e$ estimates from current galaxy samples for the
average $T^{\oiii}(Z_Q)$ and $T^{\oii}(T^{\oiii})$
relationships. This brings an empirical element into our otherwise first-principles approach. In the future, it may be possible to model the  gas temperature from first principles as well and avoid any empirical calibration steps. The fact that the thermal equilibrium temperature calculations are close to the empirical fits (see Figure~\ref{fig:TZz}) is encouraging in this regard, but we leave further exploration here to future work. 

\subsection{Comparison among the different methods}

As mentioned in the Introduction, strong line diagnostic relationships have traditionally been calibrated to direct $T_e$ measurements, to spectral synthesis models, or some combination of the two. Our work instead derived first-principles relationships for the strong line ratios. 

In Section~\ref{sec:Obs} we compared our method and traditional strong line diagnostics. The functional forms proposed here for these ratios are physically-motivated and so avoid the somewhat arbitrary polynomial fitting formula of past work.  Further, our technique clarifies the role of VCFs in determining the strong line ratios and shows how the VCFs can be extracted from O3O2. This finding helps to explain the scatter in strong line diagnostic measurements and their redshift evolution. 
Through mildly adjusting the gas temperature, strong line ratios predicted by our model can exceed the maximal values predicted in previous fitting formulas, particularly in R2 at low redshift (see Figure~\ref{fig:R}), and this allows our approach to extract reliable metallicities for more galaxies than previously possible.

Although the accuracy of our calculations has been cross-checked against \textsc{Cloudy} models, it avoids some shortcomings of these and related models. First, spectral synthesis calculations are usually limited to simple geometries, such as that of plane-parallel slabs or spherical clouds. 
Further, a single effective source of radiation is usually considered. This `one-zone' setup can lead to overestimates of $\mathrm{VCF}_\oiii$ and underestimates of $\mathrm{VCF}_\oii$ (\defcitealias{2023MNRAS.tmp.2479Y}{Y23}\citetalias{2023MNRAS.tmp.2479Y}). Although our method still assumes spherical and ionization-bounded \hii\ regions, it allows for a diversity of ISM properties and radiation field strengths/spectral shapes across the \hii\ regions in each galaxy. 

\subsection{Limitations and future prospects}

Perhaps the biggest limitation of our approach is that our empirical temperature model does not account for scatter in the TZRs. This problem is particularly acute in the case of the high-$z$ galaxy samples, where the current sample includes only 20 direct $T_e$ measurements. We are nevertheless optimistic regarding the prospects of improvements here given that \textit{JWST} will collect more direct $T_e$ measurements in the near future, and that we may be able to develop direct models for the gas temperature. 
In addition to neglecting gas temperature fluctuations, our work assumes that all emitting HII regions are in the low density limit (i.e., their densities lie much below the critical densities of interest).  It will be important to consider small-scale gas temperature and density variations in future work, since neglecting these could bias metallicity estimates 
(e.g. \citealt{2005MNRAS.362..424W,2023NatAs.tmp...87C,2023Natur.618..249M}). 

Our approach does not account for dust attenuation and should therefore be applied after reddening corrections. Note that R3 is insensitive to dust attenuation given the similar wavelengths of the [\oiii]\,5008\,\AA\ and H$\beta$ emission lines. However, the R2, R23, O3O2, and N2O2 diagnostics require careful dust corrections. Finally, line-blending is another potential concern as this can lead to biases in the luminosity estimates and resulting metallicty measurements (e.g. \citealt{2017MNRAS.465.1384C}).\par

Despite of the above caveats, this work provides a new way of modeling strong line ratios that is simple, flexible, and physically motivated. Although here we focus on R2, R3, R23, and O3O2 that are mostly relevant to the recent \textit{JWST} EoR measurements, the method proposed in this work is extendable to other strong diagnostics accessible at low-$z$, including ratios among [\sii], [\nii], [\niii], and H$\alpha$ lines. We will explore the above extensions in a future work.

\section{Acknowledgement}
AL acknowledges support through NASA ATP grant 80NSSC20K0497. AJB acknowledges support from NASA under JPL
Contract Task 70-711320, ``Maximizing Science Exploitation of
Simulated Cosmological Survey Data Across Surveys''.
HL is supported by the National Key R\&D Program of China No. 2023YFB3002502, the National Natural Science Foundation of China under No. 12373006, and the China Manned Space Project.

\section*{Data availability}
The ISM emission model introduced in this work is publicly available at \url{https://github.com/Sheng-Qi-Yang/HIILines}. 




\bibliographystyle{mnras}
\bibliography{OIII}

\begin{thebibliography}{}
\makeatletter
\relax
\def\mn@urlcharsother{\let\do\@makeother \do\$\do\&\do\#\do\^\do\_\do\%\do\~}
\def\mn@doi{\begingroup\mn@urlcharsother \@ifnextchar [ {\mn@doi@}
  {\mn@doi@[]}}
\def\mn@doi@[#1]#2{\def\@tempa{#1}\ifx\@tempa\@empty \href
  {http://dx.doi.org/#2} {doi:#2}\else \href {http://dx.doi.org/#2} {#1}\fi
  \endgroup}
\def\mn@eprint#1#2{\mn@eprint@#1:#2::\@nil}
\def\mn@eprint@arXiv#1{\href {http://arxiv.org/abs/#1} {{\tt arXiv:#1}}}
\def\mn@eprint@dblp#1{\href {http://dblp.uni-trier.de/rec/bibtex/#1.xml}
  {dblp:#1}}
\def\mn@eprint@#1:#2:#3:#4\@nil{\def\@tempa {#1}\def\@tempb {#2}\def\@tempc
  {#3}\ifx \@tempc \@empty \let \@tempc \@tempb \let \@tempb \@tempa \fi \ifx
  \@tempb \@empty \def\@tempb {arXiv}\fi \@ifundefined
  {mn@eprint@\@tempb}{\@tempb:\@tempc}{\expandafter \expandafter \csname
  mn@eprint@\@tempb\endcsname \expandafter{\@tempc}}}

\bibitem[\protect\citeauthoryear{{Allende Prieto}, {Lambert}  \&
  {Asplund}}{{Allende Prieto} et~al.}{2001}]{2001ApJ...556L..63A}
{Allende Prieto} C.,  {Lambert} D.~L.,   {Asplund} M.,  2001, \mn@doi [\apjl]
  {10.1086/322874}, \href
  {https://ui.adsabs.harvard.edu/abs/2001ApJ...556L..63A} {556, L63}

\bibitem[\protect\citeauthoryear{{Bian}, {Kewley}  \& {Dopita}}{{Bian}
  et~al.}{2018}]{2018ApJ...859..175B}
{Bian} F.,  {Kewley} L.~J.,   {Dopita} M.~A.,  2018, \mn@doi [\apj]
  {10.3847/1538-4357/aabd74}, \href
  {https://ui.adsabs.harvard.edu/abs/2018ApJ...859..175B} {859, 175}

\bibitem[\protect\citeauthoryear{{Bresolin}, {Gieren}, {Kudritzki},
  {Pietrzy{\'n}ski}, {Urbaneja}  \& {Carraro}}{{Bresolin}
  et~al.}{2009}]{2009ApJ...700..309B}
{Bresolin} F.,  {Gieren} W.,  {Kudritzki} R.-P.,  {Pietrzy{\'n}ski} G.,
  {Urbaneja} M.~A.,   {Carraro} G.,  2009, \mn@doi [\apj]
  {10.1088/0004-637X/700/1/309}, \href
  {https://ui.adsabs.harvard.edu/abs/2009ApJ...700..309B} {700, 309}

\bibitem[\protect\citeauthoryear{{Cameron}, {Katz}  \& {Rey}}{{Cameron}
  et~al.}{2023}]{2023MNRAS.522L..89C}
{Cameron} A.~J.,  {Katz} H.,   {Rey} M.~P.,  2023, \mn@doi [\mnras]
  {10.1093/mnrasl/slad046}, \href
  {https://ui.adsabs.harvard.edu/abs/2023MNRAS.522L..89C} {522, L89}

\bibitem[\protect\citeauthoryear{{Campbell}, {Terlevich}  \&
  {Melnick}}{{Campbell} et~al.}{1986}]{1986MNRAS.223..811C}
{Campbell} A.,  {Terlevich} R.,   {Melnick} J.,  1986, \mn@doi [\mnras]
  {10.1093/mnras/223.4.811}, \href
  {https://ui.adsabs.harvard.edu/abs/1986MNRAS.223..811C} {223, 811}

\bibitem[\protect\citeauthoryear{{Chen} et~al.,}{{Chen}
  et~al.}{2023}]{2023NatAs.tmp...87C}
{Chen} Y.,  et~al., 2023, \mn@doi [Nature Astronomy]
  {10.1038/s41550-023-01953-7}, \href
  {https://ui.adsabs.harvard.edu/abs/2023NatAs.tmp...87C} {}

\bibitem[\protect\citeauthoryear{{Conroy} \& {Gunn}}{{Conroy} \&
  {Gunn}}{2010}]{2010ApJ...712..833C}
{Conroy} C.,  {Gunn} J.~E.,  2010, \mn@doi [\apj]
  {10.1088/0004-637X/712/2/833}, \href
  {https://ui.adsabs.harvard.edu/abs/2010ApJ...712..833C} {712, 833}

\bibitem[\protect\citeauthoryear{{Conroy}, {Gunn}  \& {White}}{{Conroy}
  et~al.}{2009}]{2009ApJ...699..486C}
{Conroy} C.,  {Gunn} J.~E.,   {White} M.,  2009, \mn@doi [\apj]
  {10.1088/0004-637X/699/1/486}, \href
  {https://ui.adsabs.harvard.edu/abs/2009ApJ...699..486C} {699, 486}

\bibitem[\protect\citeauthoryear{{Croxall} et~al.,}{{Croxall}
  et~al.}{2013}]{2013ApJ...777...96C}
{Croxall} K.~V.,  et~al., 2013, \mn@doi [\apj] {10.1088/0004-637X/777/2/96},
  \href {https://ui.adsabs.harvard.edu/abs/2013ApJ...777...96C} {777, 96}

\bibitem[\protect\citeauthoryear{{Curti}, {Cresci}, {Mannucci}, {Marconi},
  {Maiolino}  \& {Esposito}}{{Curti} et~al.}{2017}]{2017MNRAS.465.1384C}
{Curti} M.,  {Cresci} G.,  {Mannucci} F.,  {Marconi} A.,  {Maiolino} R.,
  {Esposito} S.,  2017, \mn@doi [\mnras] {10.1093/mnras/stw2766}, \href
  {https://ui.adsabs.harvard.edu/abs/2017MNRAS.465.1384C} {465, 1384}

\bibitem[\protect\citeauthoryear{{Curti} et~al.,}{{Curti}
  et~al.}{2023}]{2023MNRAS.518..425C}
{Curti} M.,  et~al., 2023, \mn@doi [\mnras] {10.1093/mnras/stac2737}, \href
  {https://ui.adsabs.harvard.edu/abs/2023MNRAS.518..425C} {518, 425}

\bibitem[\protect\citeauthoryear{{Denicol{\'o}}, {Terlevich}  \&
  {Terlevich}}{{Denicol{\'o}} et~al.}{2002}]{2002MNRAS.330...69D}
{Denicol{\'o}} G.,  {Terlevich} R.,   {Terlevich} E.,  2002, \mn@doi [\mnras]
  {10.1046/j.1365-8711.2002.05041.x}, \href
  {https://ui.adsabs.harvard.edu/abs/2002MNRAS.330...69D} {330, 69}

\bibitem[\protect\citeauthoryear{{Dopita}, {Kewley}, {Sutherland}  \&
  {Nicholls}}{{Dopita} et~al.}{2016}]{2016Ap&SS.361...61D}
{Dopita} M.~A.,  {Kewley} L.~J.,  {Sutherland} R.~S.,   {Nicholls} D.~C.,
  2016, \mn@doi [\apss] {10.1007/s10509-016-2657-8}, \href
  {https://ui.adsabs.harvard.edu/abs/2016Ap&SS.361...61D} {361, 61}

\bibitem[\protect\citeauthoryear{{Guseva}, {Izotov}, {Fricke}  \&
  {Henkel}}{{Guseva} et~al.}{2017}]{2017A&A...599A..65G}
{Guseva} N.~G.,  {Izotov} Y.~I.,  {Fricke} K.~J.,   {Henkel} C.,  2017, \mn@doi
  [\aap] {10.1051/0004-6361/201629181}, \href
  {https://ui.adsabs.harvard.edu/abs/2017A&A...599A..65G} {599, A65}

\bibitem[\protect\citeauthoryear{{Heintz} et~al.,}{{Heintz}
  et~al.}{2023}]{2023NatAs.tmp..194H}
{Heintz} K.~E.,  et~al., 2023, \mn@doi [Nature Astronomy]
  {10.1038/s41550-023-02078-7}, \href
  {https://ui.adsabs.harvard.edu/abs/2023NatAs.tmp..194H} {}

\bibitem[\protect\citeauthoryear{{Henry} et~al.,}{{Henry}
  et~al.}{2021}]{2021ApJ...919..143H}
{Henry} A.,  et~al., 2021, \mn@doi [\apj] {10.3847/1538-4357/ac1105}, \href
  {https://ui.adsabs.harvard.edu/abs/2021ApJ...919..143H} {919, 143}

\bibitem[\protect\citeauthoryear{{Hirschmann}, {Charlot}  \&
  {Somerville}}{{Hirschmann} et~al.}{2023}]{2023arXiv230503753H}
{Hirschmann} M.,  {Charlot} S.,   {Somerville} R.~S.,  2023, \mn@doi [arXiv
  e-prints] {10.48550/arXiv.2305.03753}, \href
  {https://ui.adsabs.harvard.edu/abs/2023arXiv230503753H} {p. arXiv:2305.03753}

\bibitem[\protect\citeauthoryear{{Holweger}}{{Holweger}}{2001}]{2001AIPC..598...23H}
{Holweger} H.,  2001, in {Wimmer-Schweingruber} R.~F.,  ed.,  American
  Institute of Physics Conference Series Vol. 598, Joint SOHO/ACE workshop
  ``Solar and Galactic Composition''. pp 23--30 (\mn@eprint {arXiv}
  {astro-ph/0107426}), \mn@doi{10.1063/1.1433974}

\bibitem[\protect\citeauthoryear{{Izotov}, {Stasi{\'n}ska}, {Meynet}, {Guseva}
  \& {Thuan}}{{Izotov} et~al.}{2006}]{2006A&A...448..955I}
{Izotov} Y.~I.,  {Stasi{\'n}ska} G.,  {Meynet} G.,  {Guseva} N.~G.,   {Thuan}
  T.~X.,  2006, \mn@doi [\aap] {10.1051/0004-6361:20053763}, \href
  {https://ui.adsabs.harvard.edu/abs/2006A&A...448..955I} {448, 955}

\bibitem[\protect\citeauthoryear{{Izotov}, {Thuan}  \& {Guseva}}{{Izotov}
  et~al.}{2012}]{2012A&A...546A.122I}
{Izotov} Y.~I.,  {Thuan} T.~X.,   {Guseva} N.~G.,  2012, \mn@doi [\aap]
  {10.1051/0004-6361/201219733}, \href
  {https://ui.adsabs.harvard.edu/abs/2012A&A...546A.122I} {546, A122}

\bibitem[\protect\citeauthoryear{{Izotov}, {Thuan}, {Guseva}  \&
  {Liss}}{{Izotov} et~al.}{2018}]{2018MNRAS.473.1956I}
{Izotov} Y.~I.,  {Thuan} T.~X.,  {Guseva} N.~G.,   {Liss} S.~E.,  2018, \mn@doi
  [\mnras] {10.1093/mnras/stx2478}, \href
  {https://ui.adsabs.harvard.edu/abs/2018MNRAS.473.1956I} {473, 1956}

\bibitem[\protect\citeauthoryear{{Jones} et~al.,}{{Jones}
  et~al.}{2015a}]{2015AJ....149..107J}
{Jones} T.,  et~al., 2015a, \mn@doi [\aj] {10.1088/0004-6256/149/3/107}, \href
  {https://ui.adsabs.harvard.edu/abs/2015AJ....149..107J} {149, 107}

\bibitem[\protect\citeauthoryear{{Jones}, {Martin}  \& {Cooper}}{{Jones}
  et~al.}{2015b}]{2015ApJ...813..126J}
{Jones} T.,  {Martin} C.,   {Cooper} M.~C.,  2015b, \mn@doi [\apj]
  {10.1088/0004-637X/813/2/126}, \href
  {https://ui.adsabs.harvard.edu/abs/2015ApJ...813..126J} {813, 126}

\bibitem[\protect\citeauthoryear{{Kewley} \& {Dopita}}{{Kewley} \&
  {Dopita}}{2002}]{2002ApJS..142...35K}
{Kewley} L.~J.,  {Dopita} M.~A.,  2002, \mn@doi [\apjs] {10.1086/341326}, \href
  {https://ui.adsabs.harvard.edu/abs/2002ApJS..142...35K} {142, 35}

\bibitem[\protect\citeauthoryear{{Kobulnicky} \& {Kewley}}{{Kobulnicky} \&
  {Kewley}}{2004}]{2004ApJ...617..240K}
{Kobulnicky} H.~A.,  {Kewley} L.~J.,  2004, \mn@doi [\apj] {10.1086/425299},
  \href {https://ui.adsabs.harvard.edu/abs/2004ApJ...617..240K} {617, 240}

\bibitem[\protect\citeauthoryear{{Laseter} et~al.,}{{Laseter}
  et~al.}{2023}]{2023arXiv230603120L}
{Laseter} I.~H.,  et~al., 2023, \mn@doi [arXiv e-prints]
  {10.48550/arXiv.2306.03120}, \href
  {https://ui.adsabs.harvard.edu/abs/2023arXiv230603120L} {p. arXiv:2306.03120}

\bibitem[\protect\citeauthoryear{{Lee}, {Skillman}, {Cannon}, {Jackson},
  {Gehrz}, {Polomski}  \& {Woodward}}{{Lee} et~al.}{2006}]{2006ApJ...647..970L}
{Lee} H.,  {Skillman} E.~D.,  {Cannon} J.~M.,  {Jackson} D.~C.,  {Gehrz} R.~D.,
   {Polomski} E.~F.,   {Woodward} C.~E.,  2006, \mn@doi [\apj]
  {10.1086/505573}, \href
  {https://ui.adsabs.harvard.edu/abs/2006ApJ...647..970L} {647, 970}

\bibitem[\protect\citeauthoryear{{L{\'o}pez-S{\'a}nchez}, {Dopita}, {Kewley},
  {Zahid}, {Nicholls}  \& {Scharw{\"a}chter}}{{L{\'o}pez-S{\'a}nchez}
  et~al.}{2012}]{2012MNRAS.426.2630L}
{L{\'o}pez-S{\'a}nchez} {\'A}.~R.,  {Dopita} M.~A.,  {Kewley} L.~J.,  {Zahid}
  H.~J.,  {Nicholls} D.~C.,   {Scharw{\"a}chter} J.,  2012, \mn@doi [\mnras]
  {10.1111/j.1365-2966.2012.21145.x}, \href
  {https://ui.adsabs.harvard.edu/abs/2012MNRAS.426.2630L} {426, 2630}

\bibitem[\protect\citeauthoryear{{Ma}, {Hopkins}, {Faucher-Gigu{\`e}re},
  {Zolman}, {Muratov}, {Kere{\v{s}}}  \& {Quataert}}{{Ma}
  et~al.}{2016}]{2016MNRAS.456.2140M}
{Ma} X.,  {Hopkins} P.~F.,  {Faucher-Gigu{\`e}re} C.-A.,  {Zolman} N.,
  {Muratov} A.~L.,  {Kere{\v{s}}} D.,   {Quataert} E.,  2016, \mn@doi [\mnras]
  {10.1093/mnras/stv2659}, \href
  {https://ui.adsabs.harvard.edu/abs/2016MNRAS.456.2140M} {456, 2140}

\bibitem[\protect\citeauthoryear{{Maiolino} \& {Mannucci}}{{Maiolino} \&
  {Mannucci}}{2019}]{2019A&ARv..27....3M}
{Maiolino} R.,  {Mannucci} F.,  2019, \mn@doi [\aapr]
  {10.1007/s00159-018-0112-2}, \href
  {https://ui.adsabs.harvard.edu/abs/2019A&ARv..27....3M} {27, 3}

\bibitem[\protect\citeauthoryear{{Maiolino} et~al.,}{{Maiolino}
  et~al.}{2008}]{2008A&A...488..463M}
{Maiolino} R.,  et~al., 2008, \mn@doi [\aap] {10.1051/0004-6361:200809678},
  \href {https://ui.adsabs.harvard.edu/abs/2008A&A...488..463M} {488, 463}

\bibitem[\protect\citeauthoryear{{Mannucci} et~al.,}{{Mannucci}
  et~al.}{2009}]{2009MNRAS.398.1915M}
{Mannucci} F.,  et~al., 2009, \mn@doi [\mnras]
  {10.1111/j.1365-2966.2009.15185.x}, \href
  {https://ui.adsabs.harvard.edu/abs/2009MNRAS.398.1915M} {398, 1915}

\bibitem[\protect\citeauthoryear{{Masters} et~al.,}{{Masters}
  et~al.}{2014}]{2014AAS...22322703M}
{Masters} D.~C.,  et~al., 2014, in American Astronomical Society Meeting
  Abstracts \#223. p. 227.03

\bibitem[\protect\citeauthoryear{{McGaugh}}{{McGaugh}}{1991}]{1991ApJ...380..140M}
{McGaugh} S.~S.,  1991, \mn@doi [\apj] {10.1086/170569}, \href
  {https://ui.adsabs.harvard.edu/abs/1991ApJ...380..140M} {380, 140}

\bibitem[\protect\citeauthoryear{{M{\'e}ndez-Delgado}
  et~al.,}{{M{\'e}ndez-Delgado} et~al.}{2023a}]{2023MNRAS.523.2952M}
{M{\'e}ndez-Delgado} J.~E.,  et~al., 2023a, \mn@doi [\mnras]
  {10.1093/mnras/stad1569}, \href
  {https://ui.adsabs.harvard.edu/abs/2023MNRAS.523.2952M} {523, 2952}

\bibitem[\protect\citeauthoryear{{M{\'e}ndez-Delgado}, {Esteban},
  {Garc{\'\i}a-Rojas}, {Kreckel}  \& {Peimbert}}{{M{\'e}ndez-Delgado}
  et~al.}{2023b}]{2023Natur.618..249M}
{M{\'e}ndez-Delgado} J.~E.,  {Esteban} C.,  {Garc{\'\i}a-Rojas} J.,  {Kreckel}
  K.,   {Peimbert} M.,  2023b, \mn@doi [\nat] {10.1038/s41586-023-05956-2},
  \href {https://ui.adsabs.harvard.edu/abs/2023Natur.618..249M} {618, 249}

\bibitem[\protect\citeauthoryear{{Nagao}, {Maiolino}  \& {Marconi}}{{Nagao}
  et~al.}{2006}]{2006A&A...459...85N}
{Nagao} T.,  {Maiolino} R.,   {Marconi} A.,  2006, \mn@doi [\aap]
  {10.1051/0004-6361:20065216}, \href
  {https://ui.adsabs.harvard.edu/abs/2006A&A...459...85N} {459, 85}

\bibitem[\protect\citeauthoryear{{Nagao}, {Maiolino}, {Marconi}  \&
  {Matsuhara}}{{Nagao} et~al.}{2011}]{2011A&A...526A.149N}
{Nagao} T.,  {Maiolino} R.,  {Marconi} A.,   {Matsuhara} H.,  2011, \mn@doi
  [\aap] {10.1051/0004-6361/201015471}, \href
  {https://ui.adsabs.harvard.edu/abs/2011A&A...526A.149N} {526, A149}

\bibitem[\protect\citeauthoryear{{Nakajima}, {Ouchi}, {Isobe}, {Harikane},
  {Zhang}, {Ono}, {Umeda}  \& {Oguri}}{{Nakajima}
  et~al.}{2023}]{2023arXiv230112825N}
{Nakajima} K.,  {Ouchi} M.,  {Isobe} Y.,  {Harikane} Y.,  {Zhang} Y.,  {Ono}
  Y.,  {Umeda} H.,   {Oguri} M.,  2023, \mn@doi [arXiv e-prints]
  {10.48550/arXiv.2301.12825}, \href
  {https://ui.adsabs.harvard.edu/abs/2023arXiv230112825N} {p. arXiv:2301.12825}

\bibitem[\protect\citeauthoryear{{Patr{\'\i}cio}, {Christensen}, {Rhodin},
  {Ca{\~n}ameras}  \& {Lara-L{\'o}pez}}{{Patr{\'\i}cio}
  et~al.}{2018}]{2018MNRAS.481.3520P}
{Patr{\'\i}cio} V.,  {Christensen} L.,  {Rhodin} H.,  {Ca{\~n}ameras} R.,
  {Lara-L{\'o}pez} M.~A.,  2018, \mn@doi [\mnras] {10.1093/mnras/sty2508},
  \href {https://ui.adsabs.harvard.edu/abs/2018MNRAS.481.3520P} {481, 3520}

\bibitem[\protect\citeauthoryear{{Peimbert}, {Peimbert}  \&
  {Delgado-Inglada}}{{Peimbert} et~al.}{2017}]{2017PASP..129h2001P}
{Peimbert} M.,  {Peimbert} A.,   {Delgado-Inglada} G.,  2017, \mn@doi [\pasp]
  {10.1088/1538-3873/aa72c3}, \href
  {https://ui.adsabs.harvard.edu/abs/2017PASP..129h2001P} {129, 082001}

\bibitem[\protect\citeauthoryear{{Pettini} \& {Pagel}}{{Pettini} \&
  {Pagel}}{2004}]{2004MNRAS.348L..59P}
{Pettini} M.,  {Pagel} B. E.~J.,  2004, \mn@doi [\mnras]
  {10.1111/j.1365-2966.2004.07591.x}, \href
  {https://ui.adsabs.harvard.edu/abs/2004MNRAS.348L..59P} {348, L59}

\bibitem[\protect\citeauthoryear{{Pilyugin} \& {Thuan}}{{Pilyugin} \&
  {Thuan}}{2005}]{2005ApJ...631..231P}
{Pilyugin} L.~S.,  {Thuan} T.~X.,  2005, \mn@doi [\apj] {10.1086/432408}, \href
  {https://ui.adsabs.harvard.edu/abs/2005ApJ...631..231P} {631, 231}

\bibitem[\protect\citeauthoryear{{Pilyugin}, {Mattsson}, {V{\'\i}lchez}  \&
  {Cedr{\'e}s}}{{Pilyugin} et~al.}{2009}]{2009MNRAS.398..485P}
{Pilyugin} L.~S.,  {Mattsson} L.,  {V{\'\i}lchez} J.~M.,   {Cedr{\'e}s} B.,
  2009, \mn@doi [\mnras] {10.1111/j.1365-2966.2009.15182.x}, \href
  {https://ui.adsabs.harvard.edu/abs/2009MNRAS.398..485P} {398, 485}

\bibitem[\protect\citeauthoryear{{Pilyugin}, {V{\'\i}lchez}  \&
  {Thuan}}{{Pilyugin} et~al.}{2010}]{2010ApJ...720.1738P}
{Pilyugin} L.~S.,  {V{\'\i}lchez} J.~M.,   {Thuan} T.~X.,  2010, \mn@doi [\apj]
  {10.1088/0004-637X/720/2/1738}, \href
  {https://ui.adsabs.harvard.edu/abs/2010ApJ...720.1738P} {720, 1738}

\bibitem[\protect\citeauthoryear{{Pilyugin}, {Grebel}  \& {Kniazev}}{{Pilyugin}
  et~al.}{2014}]{2014AJ....147..131P}
{Pilyugin} L.~S.,  {Grebel} E.~K.,   {Kniazev} A.~Y.,  2014, \mn@doi [\aj]
  {10.1088/0004-6256/147/6/131}, \href
  {https://ui.adsabs.harvard.edu/abs/2014AJ....147..131P} {147, 131}

\bibitem[\protect\citeauthoryear{{Sanders} et~al.,}{{Sanders}
  et~al.}{2016}]{2016ApJ...816...23S}
{Sanders} R.~L.,  et~al., 2016, \mn@doi [\apj] {10.3847/0004-637X/816/1/23},
  \href {https://ui.adsabs.harvard.edu/abs/2016ApJ...816...23S} {816, 23}

\bibitem[\protect\citeauthoryear{{Sanders}, {Shapley}, {Topping}, {Reddy}  \&
  {Brammer}}{{Sanders} et~al.}{2024}]{2024ApJ...962...24S}
{Sanders} R.~L.,  {Shapley} A.~E.,  {Topping} M.~W.,  {Reddy} N.~A.,
  {Brammer} G.~B.,  2024, \mn@doi [\apj] {10.3847/1538-4357/ad15fc}, \href
  {https://ui.adsabs.harvard.edu/abs/2024ApJ...962...24S} {962, 24}

\bibitem[\protect\citeauthoryear{{Shen}, {Vogelsberger}, {Boylan-Kolchin},
  {Tacchella}  \& {Kannan}}{{Shen} et~al.}{2023}]{2023MNRAS.525.3254S}
{Shen} X.,  {Vogelsberger} M.,  {Boylan-Kolchin} M.,  {Tacchella} S.,
  {Kannan} R.,  2023, \mn@doi [\mnras] {10.1093/mnras/stad2508}, \href
  {https://ui.adsabs.harvard.edu/abs/2023MNRAS.525.3254S} {525, 3254}

\bibitem[\protect\citeauthoryear{{Simons} et~al.,}{{Simons}
  et~al.}{2021}]{2021ApJ...923..203S}
{Simons} R.~C.,  et~al., 2021, \mn@doi [\apj] {10.3847/1538-4357/ac28f4}, \href
  {https://ui.adsabs.harvard.edu/abs/2021ApJ...923..203S} {923, 203}

\bibitem[\protect\citeauthoryear{{Stanway}}{{Stanway}}{2020}]{2020Galax...8....6S}
{Stanway} E.~R.,  2020, \mn@doi [Galaxies] {10.3390/galaxies8010006}, \href
  {https://ui.adsabs.harvard.edu/abs/2020Galax...8....6S} {8, 6}

\bibitem[\protect\citeauthoryear{{Steidel}, {Strom}, {Pettini}, {Rudie},
  {Reddy}  \& {Trainor}}{{Steidel} et~al.}{2016}]{2016ApJ...826..159S}
{Steidel} C.~C.,  {Strom} A.~L.,  {Pettini} M.,  {Rudie} G.~C.,  {Reddy} N.~A.,
    {Trainor} R.~F.,  2016, \mn@doi [\apj] {10.3847/0004-637X/826/2/159}, \href
  {https://ui.adsabs.harvard.edu/abs/2016ApJ...826..159S} {826, 159}

\bibitem[\protect\citeauthoryear{{Sugahara}, {Inoue}, {Fudamoto}, {Hashimoto},
  {Harikane}  \& {Yamanaka}}{{Sugahara} et~al.}{2022}]{2022ApJ...935..119S}
{Sugahara} Y.,  {Inoue} A.~K.,  {Fudamoto} Y.,  {Hashimoto} T.,  {Harikane} Y.,
    {Yamanaka} S.,  2022, \mn@doi [\apj] {10.3847/1538-4357/ac7fed}, \href
  {https://ui.adsabs.harvard.edu/abs/2022ApJ...935..119S} {935, 119}

\bibitem[\protect\citeauthoryear{{Sun}, {Faucher-Gigu{\`e}re}, {Hayward},
  {Shen}, {Wetzel}  \& {Cochrane}}{{Sun} et~al.}{2023}]{2023ApJ...955L..35S}
{Sun} G.,  {Faucher-Gigu{\`e}re} C.-A.,  {Hayward} C.~C.,  {Shen} X.,  {Wetzel}
  A.,   {Cochrane} R.~K.,  2023, \mn@doi [\apjl] {10.3847/2041-8213/acf85a},
  \href {https://ui.adsabs.harvard.edu/abs/2023ApJ...955L..35S} {955, L35}

\bibitem[\protect\citeauthoryear{{Tremonti} et~al.,}{{Tremonti}
  et~al.}{2004}]{2004ApJ...613..898T}
{Tremonti} C.~A.,  et~al., 2004, \mn@doi [\apj] {10.1086/423264}, \href
  {https://ui.adsabs.harvard.edu/abs/2004ApJ...613..898T} {613, 898}

\bibitem[\protect\citeauthoryear{{Ucci} et~al.,}{{Ucci}
  et~al.}{2023}]{2023MNRAS.518.3557U}
{Ucci} G.,  et~al., 2023, \mn@doi [\mnras] {10.1093/mnras/stac2654}, \href
  {https://ui.adsabs.harvard.edu/abs/2023MNRAS.518.3557U} {518, 3557}

\bibitem[\protect\citeauthoryear{{Venturi} et~al.,}{{Venturi}
  et~al.}{2024}]{2024arXiv240303977V}
{Venturi} G.,  et~al., 2024, \mn@doi [arXiv e-prints]
  {10.48550/arXiv.2403.03977}, \href
  {https://ui.adsabs.harvard.edu/abs/2024arXiv240303977V} {p. arXiv:2403.03977}

\bibitem[\protect\citeauthoryear{{Wesson}, {Liu}  \& {Barlow}}{{Wesson}
  et~al.}{2005}]{2005MNRAS.362..424W}
{Wesson} R.,  {Liu} X.~W.,   {Barlow} M.~J.,  2005, \mn@doi [\mnras]
  {10.1111/j.1365-2966.2005.09325.x}, \href
  {https://ui.adsabs.harvard.edu/abs/2005MNRAS.362..424W} {362, 424}

\bibitem[\protect\citeauthoryear{{Yang} \& {Lidz}}{{Yang} \&
  {Lidz}}{2020}]{2020MNRAS.499.3417Y}
{Yang} S.,  {Lidz} A.,  2020, \mn@doi [\mnras] {10.1093/mnras/staa3000}, \href
  {https://ui.adsabs.harvard.edu/abs/2020MNRAS.499.3417Y} {499, 3417}

\bibitem[\protect\citeauthoryear{{Yang}, {Lidz}, {Smith}, {Benson}  \&
  {Li}}{{Yang} et~al.}{2023}]{2023MNRAS.tmp.2479Y}
{Yang} S.,  {Lidz} A.,  {Smith} A.,  {Benson} A.,   {Li} H.,  2023, \mn@doi
  [\mnras] {10.1093/mnras/stad2571}, \href
  {https://ui.adsabs.harvard.edu/abs/2023MNRAS.tmp.2479Y} {}

\bibitem[\protect\citeauthoryear{{Yates}, {Schady}, {Chen}, {Schweyer}  \&
  {Wiseman}}{{Yates} et~al.}{2020}]{2020A&A...634A.107Y}
{Yates} R.~M.,  {Schady} P.,  {Chen} T.~W.,  {Schweyer} T.,   {Wiseman} P.,
  2020, \mn@doi [\aap] {10.1051/0004-6361/201936506}, \href
  {https://ui.adsabs.harvard.edu/abs/2020A&A...634A.107Y} {634, A107}

\bibitem[\protect\citeauthoryear{{Zahid}, {Geller}, {Kewley}, {Hwang},
  {Fabricant}  \& {Kurtz}}{{Zahid} et~al.}{2013}]{2013ApJ...771L..19Z}
{Zahid} H.~J.,  {Geller} M.~J.,  {Kewley} L.~J.,  {Hwang} H.~S.,  {Fabricant}
  D.~G.,   {Kurtz} M.~J.,  2013, \mn@doi [\apjl] {10.1088/2041-8205/771/2/L19},
  \href {https://ui.adsabs.harvard.edu/abs/2013ApJ...771L..19Z} {771, L19}

\bibitem[\protect\citeauthoryear{{Zaritsky}, {Kennicutt}  \&
  {Huchra}}{{Zaritsky} et~al.}{1994}]{1994ApJ...420...87Z}
{Zaritsky} D.,  {Kennicutt} Robert~C. J.,   {Huchra} J.~P.,  1994, \mn@doi
  [\apj] {10.1086/173544}, \href
  {https://ui.adsabs.harvard.edu/abs/1994ApJ...420...87Z} {420, 87}

\makeatother
\end{thebibliography}




\label{lastpage}
\end{document}